%% file: main.tex
\pdfoutput=1
\PassOptionsToPackage{section}{placeins}
\documentclass[]{tokenwave}
\usepackage[utf8]{inputenc}
\usepackage{url}
\usepackage{amsfonts}
\usepackage{amsmath}
\usepackage{amssymb}
\usepackage{nicefrac}
\usepackage{wrapfig}
\usepackage{float}
\usepackage{needspace}
\usepackage{pifont}
\usepackage{fontawesome}
\usepackage{xspace}
\usepackage{enumitem}
\usepackage{soul}
\usepackage{tikz}
\usetikzlibrary{arrows.meta,positioning,calc,backgrounds,fit}

\crefname{figure}{Figure}{Figures}
\Crefname{figure}{Figure}{Figures}
\crefname{table}{Table}{Tables}
\Crefname{table}{Table}{Tables}
\crefname{section}{Section}{Sections}
\Crefname{section}{Section}{Sections}
\crefname{subsection}{Section}{Sections}
\Crefname{subsection}{Section}{Sections}
\crefname{appendix}{Appendix}{Appendices}
\Crefname{appendix}{Appendix}{Appendices}
\crefname{subappendix}{Appendix}{Appendices}
\Crefname{subappendix}{Appendix}{Appendices}


\titleformat*{\paragraph}{\sffamily\bfseries}

\definecolor{cbInk}{HTML}{1F2933}
\definecolor{cbInk2}{HTML}{52606D}
\definecolor{cbGrey}{HTML}{9AA5B1}
\definecolor{cbBlue}{HTML}{3A6EA5}
\definecolor{cbBlueL}{HTML}{DCE7F1}
\definecolor{cbRust}{HTML}{C2543D}
\definecolor{cbGreen}{HTML}{4E7A46}

\input{macros}

\title{SWE-CoWork: Benchmarking Coding Agents as Digital Employees
in Organizations}

\author[]{TokenWave AI\protect\footnotemark}

\abstract{\input{content/0_Abstract}}

\date{\today}
\metadata[Website]{\url{https://tokenwave-ai.github.io/swe-cowork-page}}
\metadata[Code]{\url{https://github.com/TokenWave-AI/swe-cowork-bench}}
\metadata[Dataset]{\url{https://huggingface.co/datasets/Tokenwave-AI/swe-cowork}}
\metadata[Authors]{Full author list in \Cref{sec:appendix-contributions}}

\begin{document}

\maketitle
\footnotetext{Located in Palo Alto, United States.}

\input{content/1_Introduction}

\input{content/2_RelatedWork}

\input{content/3_Benchmark}

\input{content/4_InformationDesign}

\input{content/5_Experiments}

\input{content/6_Results}

\let\sectionFloatBarrier\FloatBarrier
\let\FloatBarrier\relax
\input{content/7_Limitations}
\let\FloatBarrier\sectionFloatBarrier

\FloatBarrier

\input{main.bbl}
\clearpage
\appendix
\crefalias{section}{appendix}

\input{content/D_Contributions}

\input{content/A_Appendix}

\input{content/B_UpstreamProvenance}
\input{content/C_ObservedTrajectories}

\end{document}

%% file: macros.tex
\newcommand{\ntasks}{101\xspace}
\newcommand{\nrepos}{101\xspace}
\newcommand{\nlangs}{11\xspace}
\newcommand{\ndomains}{12\xspace}
\newcommand{\nmultilang}{11\xspace}
\newcommand{\nnodes}{32\xspace}
\newcommand{\nloc}{2,955\xspace}
\newcommand{\nfiles}{60\xspace}
\newcommand{\instrwords}{206\xspace}

\newcommand{\docwords}{5,453\xspace}

\newcommand{\totalcards}{10,329\xspace}
\newcommand{\totaldocs}{5,097\xspace}
\newcommand{\totalmsgs}{40,836\xspace}
\newcommand{\totalmrs}{1,194\xspace}
\newcommand{\totalpeople}{2,626\xspace}
\newcommand{\totalfptests}{2,699\xspace}
\newcommand{\totalpptests}{562\xspace}
\newcommand{\totalnodes}{3,261\xspace}
\newcommand{\bench}{\textsc{SWE-CoWork}\xspace}

\newcommand{\provisional}[1]{#1}

\newcommand{\modelicon}[1]{%
  \makebox[1.15em][c]{\raisebox{-0.18em}{\includegraphics[height=0.95em]{figures/#1}}}\hspace{0.25em}}
\newcommand{\iconclaude}{\modelicon{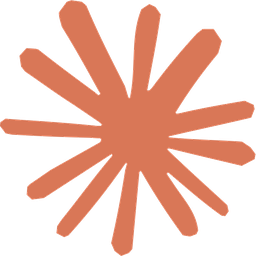}}
\newcommand{\icongpt}{\modelicon{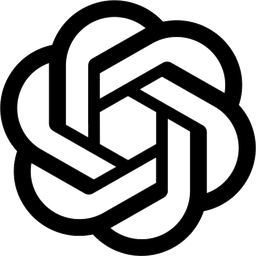}}
\newcommand{\icondeepseek}{\modelicon{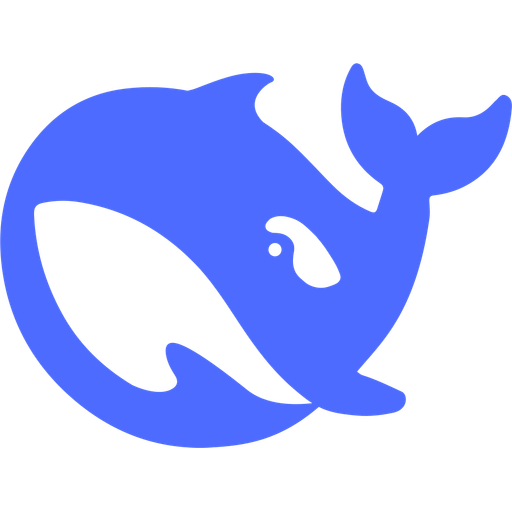}}
\newcommand{\iconglm}{\modelicon{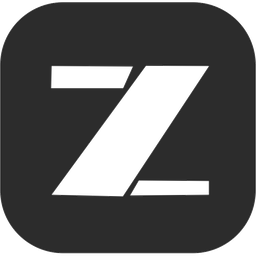}}
\newcommand{\iconkimi}{\modelicon{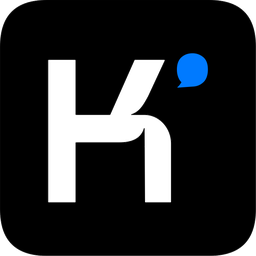}}
\newcommand{\iconqwen}{\modelicon{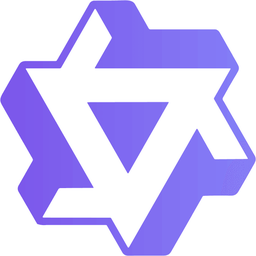}}
\newcommand{\iconseed}{\modelicon{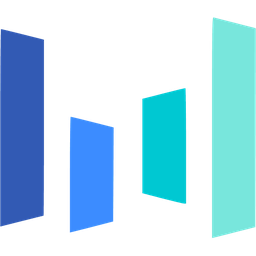}}

\definecolor{rankA}{HTML}{9FBFDE}
\definecolor{rankB}{HTML}{C5D8EB}
\definecolor{rankC}{HTML}{E3ECF5}
\newcommand{\rankfirst}[1]{\cellcolor{rankA}\textbf{#1}}
\newcommand{\ranksecond}[1]{\cellcolor{rankB}#1}
\newcommand{\rankthird}[1]{\cellcolor{rankC}#1}

%% file: content/0_Abstract.tex
A coding agent working within a software organization must establish
what to build, collaborate with colleagues, and verify what it delivers.
We introduce \bench, a benchmark comprising \ntasks development
assignments across \nrepos real repositories, \nlangs programming
languages, and \ndomains application domains. Each assignment places
an agent in a simulated engineering organization with 26 colleagues
who have distinct responsibilities, bounded knowledge, and controlled
collaboration behavior. Starting from a brief handoff, agents must
recover requirements distributed across colleagues and workplace
records, reconcile conflicting evidence, and implement changes
evaluated by private executable tests.

We evaluate eleven models and find a substantial gap between partial
implementation and complete task delivery. The two leading models
satisfy approximately \provisional{63\%} of fail-to-pass (F2P) tests
but complete fewer than \provisional{8\%} of assignments. Providing
complete requirements upfront improves F2P by approximately
\provisional{16} percentage points for Seed~Evolving and GPT-6~Astra,
yet leaves complete task success unchanged. Requirement-level
trajectory analysis reveals failures throughout the delivery process:
Claude-Opus-5.5 directs a mean of \provisional{93\%} of first questions
about queried cards to their accountable owners across runs but obtains
only \provisional{64\%} of colleague-held requirements. Deferred
questions go unfollowed, implementations remain incomplete even after
requirements are acquired, and \provisional{31\%} of requirements
reported as completed fail their tests.

These findings highlight sustained requirement tracking, from
clarification through implementation, verification, and handoff,
as a key challenge for reliable autonomous software engineering.

%% file: content/1_Introduction.tex
\section{Introduction}

Real-world software development is inherently a collaborative human endeavor rather than an isolated coding exercise.
On one hand, requirements originate from human stakeholders and evolve continuously: as intermediate milestones are reached or runtime traces surface, engineers and clients routinely refine their assumptions and introduce follow-up demands.
On the other hand, professional software engineering cannot be accomplished from a single static specification alone; it requires close coordination across diverse roles, where asking the right clarifying questions is indispensable~\citep{ko2007information,latoza2006mental,sillito2006questions,xia2018comprehension}.
Throughout this process, developers must navigate a sprawling web of project boards, versioned design documents, chat logs, merge requests, and CI histories—artifacts that frequently disagree with one another.
A design document may describe an approach that has since been superseded; a passing CI run may correspond to an outdated revision; a colleague may know a local compatibility constraint without knowing which downstream component depends on it; and a trace uncovered during implementation may invalidate an initial diagnosis.
For decades, software engineers have borne the heavy cognitive burden of checking evidence provenance, reconciling conflicting accounts, and aligning shifting requirements across the organization.

Recent advances in large language model (LLM) coding agents have achieved remarkable success on software engineering tasks, driving standard benchmarks from single-digit scores to near saturation within two years~\citep{jimenez2024swebench,anthropic2026recursive}.
In frontier industrial workflows, AI now accounts for the vast majority of code production: Anthropic reports that as of May 2026, over 80\% of the code merged into its internal codebase was authored by Claude; in the second quarter of 2026, the typical engineer merged $8\times$ as many lines of code per day as in 2024~\citep{anthropic2026recursive}.
Yet scaling raw code generation does not resolve the broader organizational challenges of software delivery.
Existing coding benchmarks~\citep{jimenez2024swebench,deepswe2026,sweatlas2026,frontierswe2026,cursorbench2026,specificrealswe2026} typically evaluate isolated, pre-scoped task descriptions. Studies of underspecification vary the information in a fixed initial request~\citep{realswe2026} or let agents query a designated simulated user who holds the missing requirements~\citep{pu2026lhaw,askorassume2026}.
Consequently, today's coding agents largely function as passive \emph{``spec executors''}: they accelerate low-level implementation, but still require humans to curate organizational context, resolve conflicting historical records, and translate evolving stakeholder needs into self-contained specifications.

To further reduce human involvement in low-level execution and elevate human engineers to high-level direction setting, prioritization, and final acceptance, coding agents must evolve from passive \emph{``spec executors''} into true \emph{digital software engineers} possessing four core organizational capabilities:
\begin{itemize}[leftmargin=1.2em,itemsep=2pt,topsep=2pt]
\item \textbf{Continuous:} Interacting with colleagues across familiar workplace interfaces (e.g., IM chat, project boards, and code reviews) while maintaining coherent context across multiple sources and evolving revisions—such as adapting execution when an early draft proposal is superseded in a later document version or when new runtime evidence arrives mid-implementation.
\item \textbf{Proactive:} Operating without step-by-step human prompting from a brief handoff (e.g., ``finish the 2026.09 upgrade; the previous task owner is on leave''), actively identifying and consulting accountable owners with concrete code context, while exercising restraint against scope distractors such as canceled cards or neighboring projects.
\item \textbf{Efficient:} Communicating with colleagues autonomously within a bounded working budget and advancing multiple complex requirements in parallel—continuing productive work on other modules rather than blocking when a colleague requests a reproduction or defers an answer.
\item \textbf{Persistent:} Remaining embedded in a realistic workplace environment over long execution horizons and context compactions, continuously tracking the state of every requirement from initial clarification through implementation, verification, and an accurate final handoff report.
\end{itemize}

\begin{figure}[!htbp]
\centering
\includegraphics[width=\textwidth]{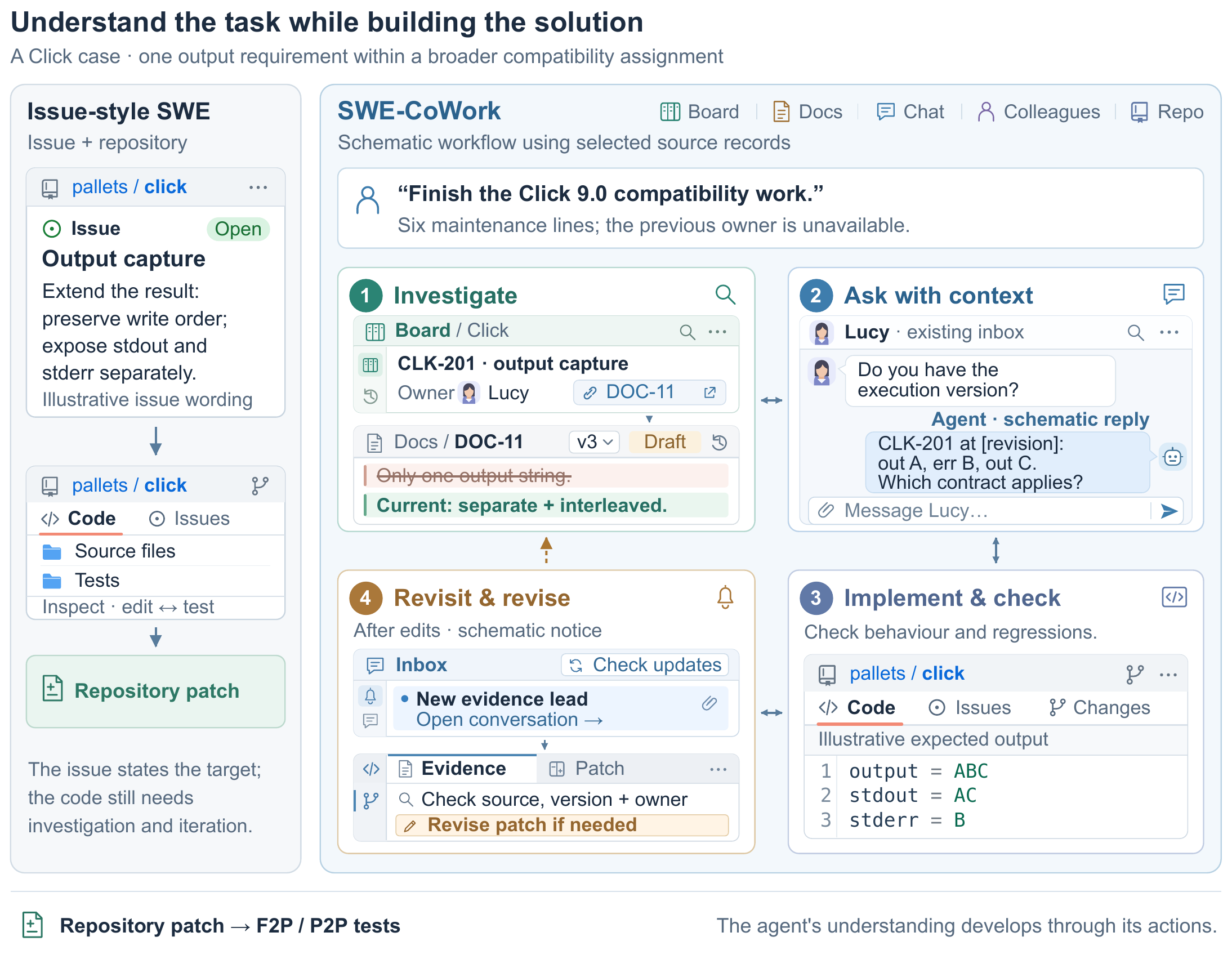}
\caption{The same Click requirement presented directly as an issue-based task
(left) and within an ongoing \bench assignment (right). In \bench, knowledge
is distributed among digital employees and workplace records. The agent
recovers the requirement, consults the responsible colleague, implements and
verifies the change, and updates its interpretation when new evidence appears.
Interfaces are schematic; Appendix~\ref{sec:appendix-cases} provides the source
records.}
\label{fig:case-contrast}
\end{figure}

Because current evaluation paradigms lack tasks that measure these organizational capabilities end to end, we introduce \bench to benchmark coding agents as digital employees inside software organizations.
Each \bench task combines a real open-source repository with an authored workplace populated by 26 \emph{digital employees}—rule-governed, LLM-backed models of colleagues with distinct responsibilities, bounded local knowledge, and controlled collaboration behavior (e.g., disclosing a held requirement only when given sufficient code context, requesting a reproduction, or referring the agent to another owner), yielding \totalpeople employee profiles across the benchmark.
Figure~\ref{fig:case-contrast} contrasts a single Click requirement presented directly as an issue against the same requirement embedded within an ongoing \bench assignment, while a typical full assignment distributes requirements across a median of six accountable owners (such as module maintainers, compatibility liaisons, and release coordinators).
In total, \bench comprises \ntasks cumulative development assignments across \nrepos upstream repositories, \nlangs languages, and \ndomains application domains, where a median task adds \nloc production lines across \nfiles files and is graded by \nnodes executable verifier \emph{test nodes}.
Starting from a short onboarding note, a frozen repository, and a workplace containing project boards, versioned documents, merge-request histories, CI records, and \totalmsgs chat messages across the benchmark (Figure~\ref{fig:engineering-challenges}), the agent must inspect records, consult colleagues, reconcile conflicting evidence, implement and verify changes, and deliver a final handoff document.

Each scored requirement is linked to an accountable owner, a discovery route through workplace records or colleague interactions, and private executable F2P/P2P tests. This linkage allows us to trace requirements from initial discovery to verified delivery. We further introduce a paired Oracle-Spec condition that provides complete requirements upfront while holding the repository, tests, scaffold, and execution budget constant. Together with agent trajectories, this comparison helps distinguish failures in requirement acquisition from those that persist during implementation, verification, and handoff.

We evaluate eleven models on \provisional{101} tasks and report F2P, P2P, and complete task pass rates (Table~\ref{tab:partial-models}).
Claude-Opus-5.5 and GPT-6~Astra pass \provisional{63.84\%} and \provisional{62.91\%} of F2P test nodes but complete only \provisional{6.93\%} and \provisional{7.92\%} of tasks; the remaining nine models reach at most \provisional{39.09\%} F2P and \provisional{0.99\%} complete task success.
Supplying complete requirements upfront in \textsc{Oracle-Spec} raises pooled F2P from \provisional{39.09\%} to \provisional{55.76\%} for Seed~Evolving and from \provisional{62.91\%} to \provisional{78.84\%} for GPT-6~Astra (Table~\ref{tab:seed-oracle-comparison}), yet complete task pass rates remain unchanged (\provisional{0.99\%} and \provisional{7.92\%}).
Requirement-level trajectory analysis reveals why: weaker models remain trapped in passive spec-executor behavior (e.g., DeepSeek~V4~Pro sends a median of \provisional{0} colleague messages per run and obtains only \provisional{1.8\%} of colleague-held requirements), whereas frontier models find the right people but fail at persistent requirement tracking.
\provisional{Specifically, Claude-Opus-5.5 directs 93\% of first questions to the accountable owner and obtains 64\% of colleague-held requirements, with only 3\% belonging to uncontacted owners.
However, failures persist after initial contact: relevant cards go unraised with already-contacted owners, deferred questions go unfollowed (153 of 450 deferred threads are abandoned with a median of 7.1 hours of budget remaining), disclosed requirements still account for 56\% of failed linked tests, and 31\% of requirements reported as ``done'' in final handoffs fail their tests.}

In summary, we make the following contributions:
\begin{itemize}[leftmargin=1.2em,itemsep=2pt,topsep=2pt]
\item \textbf{The first benchmark for organizational software engineering intelligence.}
\bench provides \ntasks development assignments across \nrepos repositories, \nlangs languages, and \ndomains domains, shifting evaluation from isolated spec execution to autonomous collaboration inside software organizations. In each assignment, requirements are distributed across rule-governed colleagues with bounded responsibilities and evolving project records, with every requirement tied to a fixed F2P/P2P test contract.

\item \textbf{A paired evaluation of organizational overhead and implementation bottlenecks.}
The \textsc{CoWork} and \textsc{Oracle-Spec} conditions share the repository, tests, scaffold, and execution budget, allowing us to jointly examine the organizational overhead of establishing requirements through colleagues and records alongside downstream implementation bottlenecks. We report this comparison for Seed~Evolving and GPT-6~Astra together with \textsc{CoWork} results across eleven models.

\item \textbf{A requirement-level account of where sustained tracking fails.}
By tracing each requirement across colleague conversations, code changes, verification, and handoff notes to individual tests, we show that weaker models struggle to proactively obtain requirements from the organization. \provisional{Frontier models such as Claude-Opus-5.5 reliably reach the right owners, yet fail at sustained requirement tracking at later stages---through unraised cards, abandoned deferrals, unsatisfied disclosed requirements, premature stops, and completion claims contradicted by tests.}
\end{itemize}

%% file: content/2_RelatedWork.tex
\section{Related Work}

\begin{table}[!htbp]
\centering
\input{tables/table_recent}
\caption{How recent software engineering benchmarks present engineering work to agents.
\emph{Initial information} describes what the agent receives at the outset;
\emph{information sources} identifies where task-relevant evidence resides;
and \emph{information process} describes how agents clarify requirements,
reconcile evidence, and respond to information encountered during execution.
``n/r'' indicates that the property is not described in the cited public material.}
\label{tab:recent}
\end{table}

\paragraph{Engineering tasks in real repositories.}
Recent software engineering benchmarks evaluate a growing range of development
activities in realistic repositories. DeepSWE
\citep{deepswe2026,deepswev11} evaluates original tasks in real repositories;
SWE Atlas \citep{sweatlas2026} covers codebase questions, test writing,
and refactoring; FrontierSWE v2
\citep{frontierswe2026} examines implementation, optimization, and research
tasks under long interaction budgets; CursorBench 4.0
\citep{cursorbench2026,cursorbenchmethod2026} derives tasks from real
developer activity; and Real-SWE \citep{specificrealswe2026}
extends evaluation to private production codebases and business tooling.
These benchmarks expand the scope, complexity, and execution horizon of
coding tasks, but generally provide agents with an identified development
objective. In \bench, agents instead join an ongoing project with a brief
handoff and must establish its current requirements from organizational
records, colleagues, and repository history before implementing and
verifying the changes.

\paragraph{Clarification and requirement recovery.}
Several benchmarks investigate how coding agents handle incomplete or
ambiguous requirements. RealSWE \citep{realswe2026} varies the information
provided for a fixed task, while LHAW \citep{pu2026lhaw} systematically
removes goals, constraints, inputs, and contextual information.
\emph{Ask or Assume} \citep{askorassume2026} and Dialogue SWE-Bench
\citep{dialogueswe2026} introduce interactive users who answer clarification
questions, and ICAE-Bench \citep{icaebench2026} grounds user responses
in withheld product constraints. $\tau^{\tau}$-Bench
\citep{shi2026hypertau} requires developer agents to recover requirements
through business artifacts and client interactions. These works demonstrate
the importance of information acquisition, typically within a task whose
scope is predefined or whose missing requirements can be clarified
through a designated user or client.

\bench examines requirement recovery within a broader engineering
organization. Requirements are distributed among multiple colleagues
with distinct responsibilities and bounded knowledge, as well as project
boards, versioned documents, conversations, reviews, and execution records.
Agents must identify the appropriate owners, ask questions with sufficient
context, follow up on deferred answers, and reconcile information from
different sources. Historical records may contain superseded decisions or
observations tied to earlier revisions, while new evidence encountered
during implementation may require revisiting an existing solution.
These interactions require agents to establish both the content and
the authority of requirements as the work progresses.

\paragraph{Agents in workplace environments.}
Workplace-agent benchmarks evaluate professional tasks involving
organizational records, applications, and collaboration with colleagues.
TheAgentCompany \citep{xu2024agentcompany} places agents in a simulated
company with coworkers and workplace services to evaluate their ability
to complete consequential professional tasks.
Recent system efforts also explore more autonomous software development
workflows. DeepSeek's Super-agent prototype
\citep{deepseeksuperagent2026}, for example, connects coding agents to
GitHub and team messaging, using event-driven task dispatch and
dedicated execution sessions to support ongoing engineering work.

\bench provides an evaluation environment for coding agents operating
within such organizational contexts. Each task combines a real software
repository with an authored engineering workplace populated by digital
employees whose knowledge and responsibilities are grounded in the
project. Agents must collaborate with these employees, recover relevant
decisions from organizational records, and implement the required changes.
Each requirement is linked to a fixed executable test contract, allowing
evaluation of both requirement acquisition and verified delivery.

The reliability of simulated colleagues is also important for interactive
evaluation. Evaluation outcomes can depend on the user simulator, and
simulated interactions may differ systematically from interactions with
human users \citep{lostinsimulation2026}.
In \bench, fixed knowledge boundaries and disclosure rules constrain
colleague interactions across runs
(Section~\ref{sec:employee-twins}).

\paragraph{Analyzing agent behavior and failures.}
Prior work also investigates the processes through which coding agents
complete engineering tasks. SWE-RPG \citep{swerpg2026} provides
representations for studying planning and clarification in software agent
trajectories, while $\tau$-bench \citep{yao2024taubench} examines
reliability across repeated interactive executions.
Coding-agent frameworks such as SWE-agent, OpenHands, and Agentless
\citep{yang2024sweagent,wang2025openhands,xia2024agentless}
further demonstrate the influence of agent scaffolding on execution
behavior and task outcomes.

\bench enables requirement-level analysis of the full engineering
workflow. By linking individual requirements to colleague interactions,
workplace evidence, executable tests, and final delivery reports, we
can identify where requirements are lost between discovery, clarification,
implementation, verification, and handoff. This distinguishes failures
to acquire necessary information from failures to implement requirements
that the agent has already obtained, and reveals discrepancies between
reported completion and verified outcomes.

Table~\ref{tab:recent} compares recent benchmarks by their initial
task information, available evidence sources, and information
processes. Appendix~\ref{sec:appendix-related} discusses additional
repository, interactive, and workplace benchmarks.

%% file: tables/table_recent.tex
%

\begingroup

\small
\setlength{\tabcolsep}{3pt}
\renewcommand{\arraystretch}{1.28}
\newlength{\recentContentWidth}
\setlength{\recentContentWidth}{\dimexpr\linewidth-12\tabcolsep\relax}


\definecolor{tagTaskBG}{HTML}{EEF1F4}
\definecolor{tagTaskFG}{HTML}{46515C}

\definecolor{tagSpecBG}{HTML}{E8F1FB}
\definecolor{tagSpecFG}{HTML}{315F8C}

\definecolor{tagSourceBG}{HTML}{E6F2EE}
\definecolor{tagSourceFG}{HTML}{27685A}

\definecolor{tagProcessBG}{HTML}{F0EAF6}
\definecolor{tagProcessFG}{HTML}{705090}

\definecolor{tagEvalBG}{HTML}{F8EEDC}
\definecolor{tagEvalFG}{HTML}{8A5918}

\definecolor{tagNRBG}{HTML}{F2F2F2}
\definecolor{tagNRFG}{HTML}{777777}


\newcommand{\recentTag}[3]{%
  \begingroup
  \setlength{\fboxsep}{1.4pt}%
  \colorbox{#1}{\textcolor{#2}{\scriptsize\strut #3}}%
  \endgroup%
}

\newcommand{\tagTask}[1]{%
  \recentTag{tagTaskBG}{tagTaskFG}{#1}}

\newcommand{\tagSpec}[1]{%
  \recentTag{tagSpecBG}{tagSpecFG}{#1}}

\newcommand{\tagSource}[1]{%
  \recentTag{tagSourceBG}{tagSourceFG}{#1}}

\newcommand{\tagProcess}[1]{%
  \recentTag{tagProcessBG}{tagProcessFG}{#1}}

\newcommand{\tagEval}[1]{%
  \recentTag{tagEvalBG}{tagEvalFG}{#1}}

\newcommand{\tagNR}{%
  \recentTag{tagNRBG}{tagNRFG}{n/r}}


\begin{tabular}{
>{\raggedright\arraybackslash}p{0.190\recentContentWidth}
>{\raggedright\arraybackslash}p{0.160\recentContentWidth}
>{\raggedright\arraybackslash}p{0.145\recentContentWidth}
>{\raggedright\arraybackslash}p{0.130\recentContentWidth}
>{\raggedright\arraybackslash}p{0.225\recentContentWidth}
>{\raggedright\arraybackslash}p{0.150\recentContentWidth}
}

\toprule

\textbf{Benchmark}
&
\textbf{Engineering}\newline\textbf{task}
&
\textbf{Initial}\newline\textbf{information}
&
\textbf{Information}\newline\textbf{sources}
&
\textbf{Information}\newline\textbf{process}
&
\textbf{Evaluation}
\\

\midrule


\textbf{DeepSWE 1.1}\newline
\citep{deepswe2026,deepswev11}
&
\tagTask{Repo change}
&
\tagSpec{Request}
\tagSpec{Underspecified}
&
\tagSource{Repo}
&
\tagNR
&
\tagEval{Patch tests}
\\[7pt]


\textbf{SWE Atlas}\newline
\citep{sweatlas2026}
&
\tagTask{Code Q\&A}
\tagTask{Test writing}
\tagTask{Refactor}
&
\tagSpec{Request}
\tagSpec{Underspecified}
&
\tagSource{Repo}
&
\tagProcess{No live Q\&A}
&
\tagEval{Execution}
\tagEval{Rubric}
\\[7pt]


\textbf{FrontierSWE v2}\newline
\citep{frontierswe2026}
&
\tagTask{Build}
\tagTask{Optimize}
\tagTask{Research}
&
\tagSpec{Brief}
&
\tagSource{Workspace}
\tagSource{Self-check}
&
\tagNR
&
\tagEval{Task score}
\\[7pt]


\textbf{CursorBench 4.0}\newline
\citep{cursorbench2026,cursorbenchmethod2026}
&
\tagTask{Edit}
\tagTask{Refactor}
\tagTask{Investigate}
&
\tagSpec{Request}
\tagSpec{Underspecified}
&
\tagSource{Code}
\tagSource{Tools}
&
\tagNR
&
\tagEval{Agentic grader}
\\[7pt]


\textbf{Real-SWE}\newline
(Specific Labs)\newline
\citep{specificrealswe2026}
&
\tagTask{Enterprise change}
&
\tagSpec{Brief}
\tagSpec{Underspecified}
&
\tagSource{Repo}
\tagSource{Docs}
\tagSource{Board}
\tagSource{Chat}
&
\tagNR
&
\tagNR
\\[7pt]


\textbf{RealSWE}\newline
(paper v2)\newline
\citep{realswe2026}
&
\tagTask{Bug fix}
\tagTask{Feature}
&
\tagSpec{Issue variants}
\tagSpec{Underspecified}
&
\tagSource{Repo}
&
\tagProcess{Static request}
\tagProcess{No live Q\&A}
&
\tagEval{Fixed target}
\\[7pt]


\textbf{$\tau^{\tau}$-Bench}\newline
\citep{shi2026hypertau}
&
\tagTask{Agent build}
&
\tagSpec{Underspecified}
&
\tagSource{Docs}
\tagSource{APIs}
\tagSource{Code}
&
\tagProcess{Client Q\&A}\newline
\tagProcess{Conflicting history}\newline
\tagProcess{Superseded records}
&
\tagEval{User simulation}
\\

\midrule


\rowcolor{cbBlueL!35}
\textbf{\bench}\newline
(ours)
&
\tagTask{Multi-PR}
\tagTask{Repo change}
&
\tagSpec{Onboarding brief}
\tagSpec{Underspecified}
&
\tagSource{Repo}
\tagSource{Board}
\tagSource{Docs}
\tagSource{Chat}
\tagSource{MR/CI}
&
\tagProcess{Multi-owner Q\&A}\newline
\tagProcess{Versioned records}\newline
\tagProcess{Conflicting history}\newline
\tagProcess{Execution-time leads}
&
\tagEval{F2P/P2P}
\tagEval{Fixed contract}
\\

\bottomrule

\end{tabular}

\par\vspace{4pt}

{\footnotesize\raggedright
\tagTask{task}
\;
\tagSpec{initial information}
\;
\tagSource{information source}
\;
\tagProcess{interaction / dynamics}
\;
\tagEval{evaluation}
\par\vspace{2pt}
\tagNR\ = not reported in the cited public description.
\par
}

\endgroup

%% file: content/3_Benchmark.tex
\section{Engineering Organizations of Digital Employees}
\label{sec:benchmark}

Each \bench task places a coding agent in an engineering
organization consisting of a real software repository,
a shared workplace, and a population of digital employees.
Starting from a brief handoff, the agent establishes requirements
through workplace records and interactions with colleagues,
then implements, verifies, and delivers the requested changes.
To evaluate this process, each task defines a fixed executable
contract through fail-to-pass (F2P) and pass-to-pass (P2P) tests.
For every scored requirement, the design connects the colleague
or record that holds it, the responsibility boundary that
determines who can answer for it, and the test nodes that check it.
These links let us trace a requirement from the first question
about it through the delivered code to the final handoff note.

\subsection{Digital Employees}
\label{sec:employee-twins}

We construct an organization of 26 modeled colleagues
around each repository, yielding \totalpeople employee
profiles across the \ntasks tasks. Each digital employee represents
a named employee with a defined role, responsibilities,
local knowledge, and rules governing interactions with
the agent, with responsibilities and knowledge grounded
in the repository's files, declarations, and modules.
The organization spans one target project and four neighboring
projects that provide additional context; within this population,
a median of six employees hold requirements that determine the
task's executable outcome (Table~\ref{tab:stats}).

\paragraph{What a digital employee represents.}
Each digital employee has three properties: responsibility, local
knowledge, and collaboration behavior. \emph{Responsibility}
identifies the employee accountable for a module or decision.
\emph{Local knowledge} defines which requirements the
employee can explain and which questions require referral
to another colleague. \emph{Collaboration behavior}
governs how the employee responds. An owner may request a
reproduction or caller path, address one topic at a time,
or defer an answer until the agent follows up. The
environment controls which information can be disclosed
at each step, while a language model generates the
employee's response.

\paragraph{From individual digital employees to an organization.}
The digital employees share project records that connect people
to engineering work. Board cards identify owners,
documents record decisions, and chat and review threads
preserve earlier discussions. Information obtained from
one employee may need to be checked against another
employee's constraints or a more recent repository
revision, requiring the agent to identify appropriate owners
and resolve outstanding questions as implementation proceeds.

The agent interacts with the organization through a
dedicated tool interface. It can search the staff
directory, inspect the project board, open versioned
documents, search and read chat, message colleagues,
inspect merge requests and CI records, and check for
new activity. Digital employees are accessible through
the same workplace that contains their decisions
and work history.

\paragraph{The handoff.}
The agent joins an engineering assignment already in
progress. It begins with a short onboarding note that
states the development objective and identifies workplace
resources relevant to the current project state. The
note specifies restrictions on modifying existing tests,
excludes neighboring projects from the assignment,
and identifies historical terminology that may differ
from the current repository. The remaining acceptance
criteria are distributed across workplace records and
colleagues.

The agent decides which records to inspect, which
references to follow, and which colleagues to contact.
It interprets the available evidence against the
current repository state and resolves questions that
affect the implementation. At completion, it produces
a delivery note describing the requirements established,
the changes made, the validation performed, and any
unresolved questions.

\subsection{Composition}

\begin{figure}[!htbp]
\centering
\includegraphics[width=\linewidth]{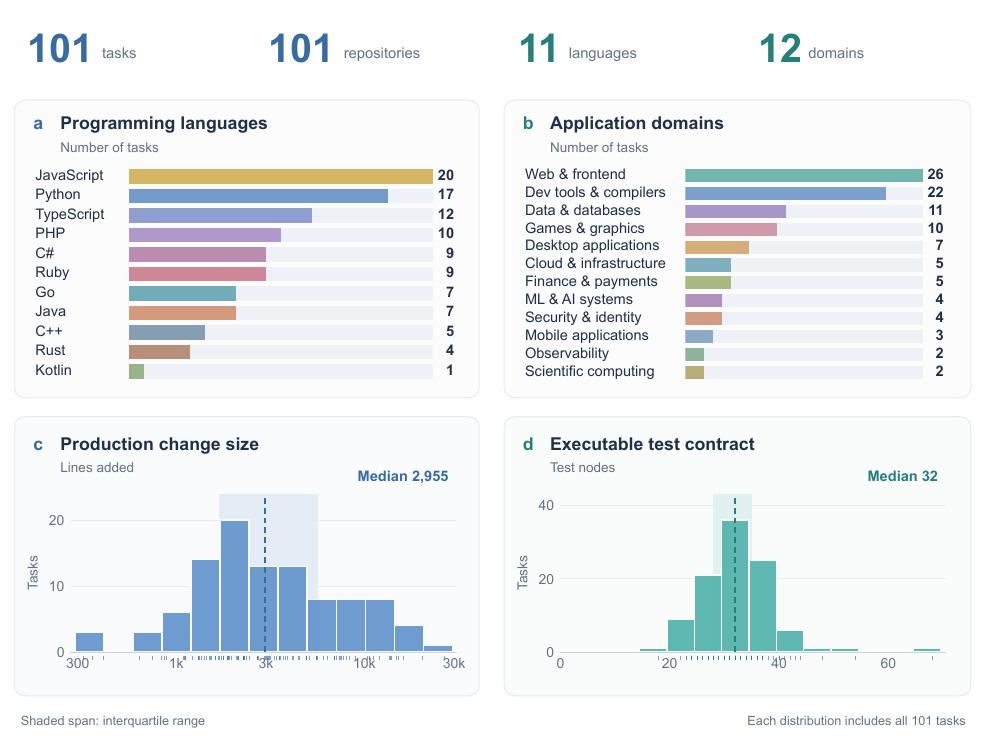}
\caption{Composition of the \ntasks \bench tasks.
(a) Primary language of the production changes.
(b) Application domain of the upstream repository.
(c) Production lines added by the reference changes,
shown on a logarithmic axis.
(d) Verifier test nodes per task.
In (c) and (d), dashed lines mark medians, shaded bands span the
interquartile range, and ticks below each axis mark individual tasks.}
\label{fig:composition}
\end{figure}

\bench contains \ntasks tasks drawn from \nrepos
upstream repositories across \nlangs programming
languages and \ndomains application domains, including
web frameworks, compilers, games, payments, and scientific
software (Figure~\ref{fig:composition};
Appendix~\ref{sec:appendix-repositories}).
\nmultilang tasks involve substantive changes in more
than one programming language.

The reference solutions involve substantial engineering
work. The median solution adds \nloc lines of production
code across \nfiles files, excluding tests, documentation,
configuration, fixtures, and generated artifacts.
For each task, we maintain a complete \textsc{Oracle-Spec}
document that enumerates the requirements encoded by
the same executable contract used to evaluate the
workplace assignment. Across the \ntasks tasks, these
contracts contain \provisional{\totalfptests} F2P test nodes
and \provisional{\totalpptests} P2P test nodes, giving
\provisional{\totalnodes} executable checks in total.

\subsection{Shared Records and Organizational History}

Each assignment is situated within the records of a
wider engineering organization. Across the benchmark
release, these organizations contain \totalcards board
cards, \totaldocs versioned documents, \totalmsgs chat
messages, \totalmrs merge requests, and \totalpeople
colleague profiles. Whereas the initial onboarding note contains
a median of \instrwords words, the authored document corpus
contains a median of \docwords words, with requirements and
supporting evidence distributed across these records and
their histories.

The agent determines which artifacts to inspect,
which historical decisions still apply, which execution
records correspond to the current revision, and which
colleagues can provide missing information. We keep
the overall organization size roughly comparable
across tasks to limit variation in the volume of
available workplace material.

\paragraph{Organizational history.}
The workplace records how projects develop over time.
Versioned documents, board histories, review threads,
CI records, and conversations contain intermediate
diagnoses, superseded plans, rejected approaches, and
observations tied to particular revisions. Historical
records may use interface, module, or protocol names
that differ from those in the current codebase. The
agent establishes which claims remain applicable
and reconciles them with subsequent decisions and the
current repository.

Access follows the organization's information boundaries.
Private groups and direct messages that do not include
the agent are inaccessible. Records from neighboring
projects provide organizational context, while those
projects remain outside the scope of the assignment.

\subsection{Scoring}

We report three outcome metrics: \textbf{F2P}, the
proportion of fail-to-pass verifier test nodes satisfied
by the submitted implementation; \textbf{P2P}, the
proportion of pass-to-pass test nodes preserved; and
\textbf{Pass rate}, the proportion of tasks for which
every F2P and P2P test node passes in the same run.
F2P and P2P are computed by pooling passed test nodes
and total test nodes across tasks.

Because each task contains multiple independently testable
behaviors, F2P captures partial delivery, while Pass rate
requires completion of the full executable contract.
Section~\ref{sec:experiments} provides the aggregation
formulas and describes the trajectory diagnostics.

%% file: content/4_InformationDesign.tex
\section{Working a \bench Task}
\label{sec:information}

In a \bench task, the agent works with the digital employees
introduced in Section~\ref{sec:employee-twins}. Starting from a brief
handoff, it establishes the requirements through repository evidence,
workplace records, and conversations with colleagues. The agent then
implements and verifies the requested changes, revisiting earlier
decisions as new information becomes available. Requirement discovery
and implementation are interleaved: an owner's clarification can change
the behavior that a patch must satisfy.

\subsection{Working with the Digital Employees}

Agents continuously move between investigating requirements, consulting colleagues, and implementing changes. Starting from a brief handoff, an agent explores the repository and workplace records to establish the current scope, identify unresolved requirements, and locate responsible colleagues. It must reconcile conflicting evidence, trace references across historical revisions, and ask questions grounded in the available code and project context. A typical run involves the following activities:

\begin{enumerate}[leftmargin=1.55em,itemsep=2pt,topsep=3pt]
\item \textbf{Orient.} Read the onboarding note, inspect the repository,
  identify the target project, and establish the initial scope and
  relevant contacts.
\item \textbf{Investigate.} Search project boards, documents, chat, and
  engineering history for requirements and implementation constraints.
  Check record provenance, reconcile version differences, and trace
  references to renamed interfaces.
\item \textbf{Ask.} Identify the colleague responsible for an unresolved
  decision and ask a question supported by the available evidence.
  Follow referrals and respond to requests for additional context.
\item \textbf{Build and verify.} Implement the required changes, run
  relevant tests, and check for regressions in existing functionality
  and neighboring projects.
\item \textbf{Revise.} Examine new messages, traces, and corrections.
  Confirm their implications with the responsible colleague and update
  the implementation when necessary.
\item \textbf{Deliver.} Document the confirmed requirements,
  implementation changes, and verification results in the delivery note.
\end{enumerate}

These activities are interleaved throughout execution. A colleague may request additional evidence, refer the agent to another owner, or introduce a constraint that requires revising an existing implementation. While some questions remain unresolved, the agent can continue working on other requirements, validate its changes, and revisit earlier decisions as new information becomes available. At handoff, it reports the implemented changes, verification results, and outstanding issues.

\subsection{Inside a Run: An Owner Clarification Changes the Patch}
\label{sec:kimai}

\begin{figure}[!htbp]
\centering
\includegraphics[width=\linewidth]{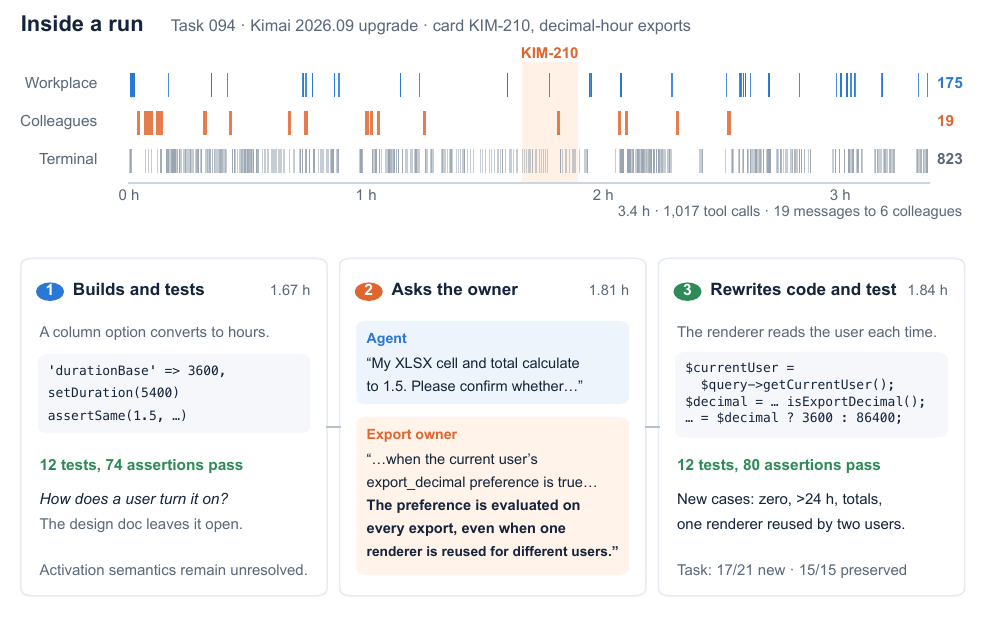}
\caption{An owner clarification changes a timesheet export implementation
in a recorded \bench run (task 094, Kimai). The agent initially implements
and tests decimal conversion, then contacts the export owner to resolve
how users activate the feature. The owner specifies that the current
user's preference must be evaluated on every render. The agent updates
the implementation and expands the tests accordingly. Times indicate
elapsed hours from the start of the run, and quotations come from the
recorded exchange.}
\label{fig:kimai}
\end{figure}

Task 094 involves a cumulative upgrade of Kimai, a PHP time-tracking
application. One requirement, KIM-210, concerns exporting timesheet
durations in decimal hours. The agent initially implemented the
numerical conversion and verified that a 5,400-second entry produced
1.5 hours. However, the design document did not specify how the
feature should be activated (Figure~\ref{fig:kimai}).

The agent then contacted the export owner to clarify the intended
behavior:

\begin{quote}\small
``My XLSX cell and total calculate to 1.5. Please confirm whether
that is the intended decimal behavior\ldots''
\hfill\emph{--- agent}
\end{quote}

\begin{quote}\small
``\ldots when the current user's
\texttt{timesheet.export\_decimal} preference is true,
the value is 1.5 hours\ldots\ The preference is evaluated
on every export, even when one renderer is reused for
different users.''
\hfill\emph{--- export owner}
\end{quote}

The clarification revealed two requirements absent from the initial
implementation: decimal formatting must follow the current user's
preference, and that preference must be evaluated on every export.
The agent revised the renderer accordingly and extended its tests
to cover consecutive exports by different users sharing the same
renderer.

This interaction illustrates how a seemingly correct implementation
can remain incomplete until a colleague clarifies the intended
behavior. Across the full assignment, the agent satisfied 17 of
21 new behaviors while preserving all 15 existing behaviors.

\FloatBarrier
\subsection{Workplace Information Challenges}
\label{sec:obstacles}

\begin{figure}[!htbp]
\centering
\includegraphics[width=\linewidth]{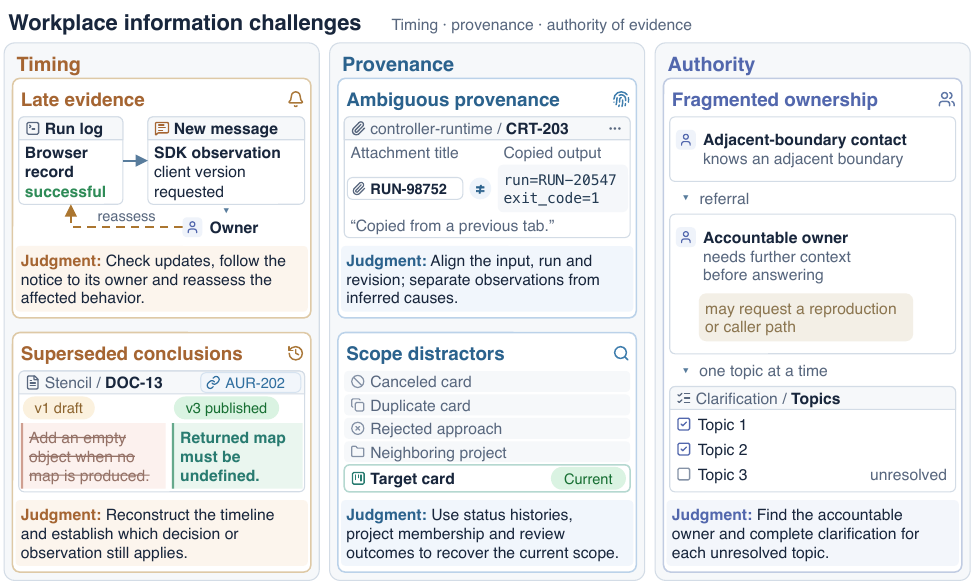}
\caption{Five workplace information challenges and the judgments they
require, organized by when evidence becomes available or is superseded,
which run or project it belongs to, and who can resolve the outstanding
decision. The panels use schematic situations and condensed
workplace-record examples. Source details for
Stencil and controller-runtime appear in
Appendix~\ref{sec:appendix-cases}.}
\label{fig:engineering-challenges}
\end{figure}

Because requirements are distributed across records and colleagues,
establishing what a record means requires checking when it became
available, where it came from, and who has authority over the decision.
Figure~\ref{fig:engineering-challenges}
groups these information challenges into five situations and shows the
engineering judgment required in each. These situations affect how
the agent establishes the requirements,
with each misleading record accompanied by an accessible correction
or a responsible colleague who can establish which evidence applies.

\paragraph{Late evidence.}
A recovered trace, corrected version label, or message from another
team may arrive after implementation begins. In the schematic example,
a successful browser record is followed by an SDK observation and a
request to confirm the client version. The agent must follow this lead
to the responsible owner, establish which behavior it concerns, and
reassess the affected code and tests. Owners remain available for
consultation before a scheduled notice appears.

\paragraph{Superseded conclusions.}
An earlier proposal can remain searchable after a published correction.
For Stencil's \texttt{AUR-202}, the draft of \texttt{DOC-13} proposes
an empty object when no source map is generated. The published third
version instead requires the returned \texttt{map} field to remain
\texttt{undefined}, while retaining the chunk's other output information.
The agent must compare revisions and publication states to identify
the applicable rule.

\Needspace{6\baselineskip}
\paragraph{Ambiguous provenance.}
A log's title, contents, and surrounding discussion may refer to different
executions. In the controller-runtime example, the attachment title
names \texttt{RUN-98752}, but the copied output identifies
\texttt{RUN-20547} and states that it came from a previous tab.
The agent must align the output with its input, run, and code revision
before using it to diagnose a failure.

\paragraph{Scope distractors.}
Canceled cards, duplicates, rejected approaches, and neighboring
projects remain visible alongside the current assignment. Similar
wording can therefore lead to a record outside the required change.
Status histories, project membership, and review outcomes establish
which card is current and which approaches have already been rejected.
The agent uses these distinctions to recover the target scope.

\paragraph{Fragmented ownership.}
A colleague may know an adjacent interface boundary yet lack authority
over the unresolved decision. A referral leads to the accountable owner,
who may request a reproduction, caller path, or version before answering.
Clarification also proceeds topic by topic: resolving one question can
leave others open. The agent supplies the requested context and tracks
which decisions each exchange establishes before carrying them into
the implementation.

These situations make investigation and coding interdependent. A newly
located record can prompt another question, whose answer changes the
patch or its tests. Appendix~\ref{sec:appendix-cases} gives the underlying
records for the Stencil and controller-runtime examples and a further
Click example involving output semantics and stream ownership.

%% file: content/5_Experiments.tex
\section{Evaluation}
\label{sec:experiments}

We evaluate eleven coding models in the \textsc{CoWork} setting
using executable outcomes and collaboration records. A paired
comparison with \textsc{Oracle-Spec} measures the performance of
Seed~Evolving and GPT-6~Astra when complete requirements are supplied
at the start. To examine how requirements are obtained, incorporated
into code, verified, and reported at handoff, we analyze
Claude-Opus-5.5 trajectories.

\subsection{Primary metrics}

We report \textbf{F2P}, \textbf{P2P}, and \textbf{Pass rate} on
$N=\provisional{101}$ tasks. For task $i$, let $f_i$ and $p_i$
denote the numbers of passing F2P and P2P test nodes, and let
$F_i$ and $P_i$ denote their respective totals. All metrics
are expressed as percentages:
\begin{align*}
\mathrm{F2P} &= 100\,\frac{\sum_{i=1}^{N} f_i}{\sum_{i=1}^{N} F_i},
&\mathrm{P2P} &= 100\,\frac{\sum_{i=1}^{N} p_i}{\sum_{i=1}^{N} P_i},\\
\mathrm{Pass\ rate} &= \frac{100}{N}\sum_{i=1}^{N}
\mathbf{1}\!\left[f_i=F_i\ \land\ p_i=P_i\right].
\end{align*}

F2P measures how many new requirements are satisfied, while P2P
measures whether existing behavior is preserved. Both metrics
aggregate test nodes across tasks, giving greater weight to tasks
with more test nodes. Pass rate measures the fraction of tasks for which
every F2P and P2P test node passes.

All models are evaluated against the full release contract:
\provisional{\totalfptests} F2P test nodes,
\provisional{\totalpptests} P2P test nodes, and
\provisional{101} tasks. Only passes confirmed by the verifier count
toward the numerators. For trajectory analysis, we use the
fraction of F2P test nodes passed within each task to relate
outcomes to agent behavior.

\subsection{Experimental design}

\paragraph{CoWork evaluation.}
We evaluate eleven models at max effort in the \textsc{CoWork}
setting: Claude-Opus-5.5~\citep{claudeopus55},
GPT-6~Astra~\citep{gpt6astra},
Seed~2.1~Pro~\citep{seed21},
ByteDance's Seed~Evolving~\citep{seedevolving},
Kimi~K3~\citep{kimik3},
DeepSeek~V4~Pro~\citep{deepseekv4pro},
DeepSeek~V4.1~Flash~\citep{deepseekv41flash},
GLM~5.3~\citep{glm53},
GLM~5.2~\citep{glm52},
and Qwen3.8-Max and Qwen3.8-Flash~\citep{qwen38}.

GPT-6~Astra uses a Codex scaffold, and the other ten models
use Claude Code scaffolds. In every \textsc{CoWork} run, the digital
employees are backed by GPT-5.6~Terra at high reasoning effort, so all
agents interact with the same colleague model. Each agent starts with a brief
handoff and access to the organization. It identifies
the requirements through workplace records and conversations
with colleagues, implements the requested changes, and submits
its work for F2P/P2P evaluation. Table~\ref{tab:partial-models}
reports F2P, P2P, and Pass rate for all eleven models on
the \provisional{101}-task release contract.

\paragraph{Oracle-Spec comparison.}
We evaluate Seed~Evolving and GPT-6~Astra under two conditions.
In \textsc{CoWork}, the agent obtains requirements through
workplace records and colleagues. In \textsc{Oracle-Spec},
the complete requirement set is provided at the start, and
the agent has no colleagues to consult.
Both conditions cover the same \provisional{101} tasks,
with identical task definitions, repository states, and
executable test contracts.

We compare F2P, P2P, and Pass rate across the paired runs
to measure how access to complete requirements affects
implementation outcomes.

\paragraph{Claude-Opus-5.5: trajectory analysis.}
We examine Claude-Opus-5.5's \textsc{CoWork} trajectories
by combining programmatic disclosure logs with a rubric-based
LLM trajectory reviewer.
For each run, we trace which requirements reached the agent,
which remained unresolved, how it followed up with colleagues,
and how it verified and reported its work.

We select Claude-Opus-5.5 because it achieves the highest
CoWork F2P score among the evaluated models. The analysis covers its recorded
runs and examines the relationship between requirement
acquisition, agent behavior, and implementation outcomes.

Supplementary diagnostics appear in Appendix~\ref{sec:appendix-results};
Appendix~\ref{sec:appendix-observed} gives the measurement definitions,
claim-classification rules, and evidence sources.

\subsection{Supplementary analysis metrics}
\label{sec:analysis-metrics}

\paragraph{Model calls and token usage.}
We report the median per-run
number of model calls, input tokens, output tokens, tool calls,
and context compactions. Input usage accumulates tokens processed
across calls within a run, including repeated context. We pair these usage
medians with each model's F2P on the full release contract.
Figures~\ref{fig:usage-medians} and~\ref{fig:usage-tools} in
Appendix~\ref{sec:appendix-usage} report these comparisons.

\paragraph{Workplace exploration and the first code edit.}
For eight models, we report per-run medians of workplace queries
before the first code edit, the tool-call index of that edit, and
workplace queries over the full run (Figure~\ref{fig:workplace-exploration}).
The same comparison includes median messages sent to colleagues.

\paragraph{Requirement acquisition and collaboration.}
Each task's \textsc{Oracle-Spec} enumerates its requirements
and identifies the responsible colleague or public record.
Disclosure logs and saved tool replies establish which
requirements were released by colleagues and which replies
reached the agent.

We measure \textbf{ask rate}, the fraction of requirements
held by colleagues that the agent raised with the responsible
owner or obtained; \textbf{targeting}, the fraction of queried
cards whose first question was directed to the accountable
owner; \textbf{gain per question}, the number of disclosures
per colleague message; \textbf{follow-through}, the fraction
of raised requirements that were subsequently disclosed;
\textbf{redundancy}, questions about cards whose requirements
had already been fully disclosed; and \textbf{unread disclosures},
replies containing disclosed requirements that were absent
from the agent's recorded observations.

%% file: content/6_Results.tex
\section{Results}
\label{sec:results}

Across eleven models, requirement acquisition closely tracks delivery,
while model-call and tool-call counts do not consistently track F2P.
Supplying complete requirements improves partial delivery for
Seed~Evolving and GPT-6~Astra without increasing complete task success.
Claude-Opus-5.5 trajectories locate further failures after initial
contact: deferred questions remain unresolved, obtained requirements
fail their tests, and handoff claims overstate completed work.

\subsection{Results Across Models}
\label{sec:partial-models}
\label{sec:observed-runs}

Table~\ref{tab:partial-models} compares eleven models on the
\provisional{101}-task CoWork benchmark. Even the leading models
rarely complete an entire assignment.

\begin{table}[!htbp]
\centering
\small
\setlength{\tabcolsep}{12pt}
\input{tables/table_partial_models}
\caption{\textbf{CoWork evaluation on \provisional{101} tasks.}
F2P and P2P use the release totals of
\provisional{\totalfptests} and
\provisional{\totalpptests} test nodes, respectively.
Pass rate requires every F2P and P2P test node to pass in the same run.
All scores are percentages, with higher values indicating better
outcomes. Shading marks the top three per metric, from darkest
(best, bold) to lightest; ties share a shade.
Rows are ordered by F2P.}
\label{tab:partial-models}
\end{table}

\paragraph{Partial delivery versus complete task success.}
Claude-Opus-5.5 and GPT-6 Astra achieve F2P scores of
63.84\% and 62.91\%, respectively, yet their complete task
pass rates are only 6.93\% and 7.92\%. The remaining nine
models achieve at most 39.09\% F2P and 0.99\% complete
task success. These results reveal a substantial gap
between implementing individual required behaviors and
satisfying the full assignment.

P2P scores are comparatively higher, ranging from
81.85\% to 95.20\%, with rankings that differ from F2P.
This suggests that preserving existing behavior and
delivering new requirements pose distinct challenges,
with the latter remaining a major bottleneck in
organizational software engineering.

\begin{figure}[!htbp]
\centering
\includegraphics[width=\linewidth]{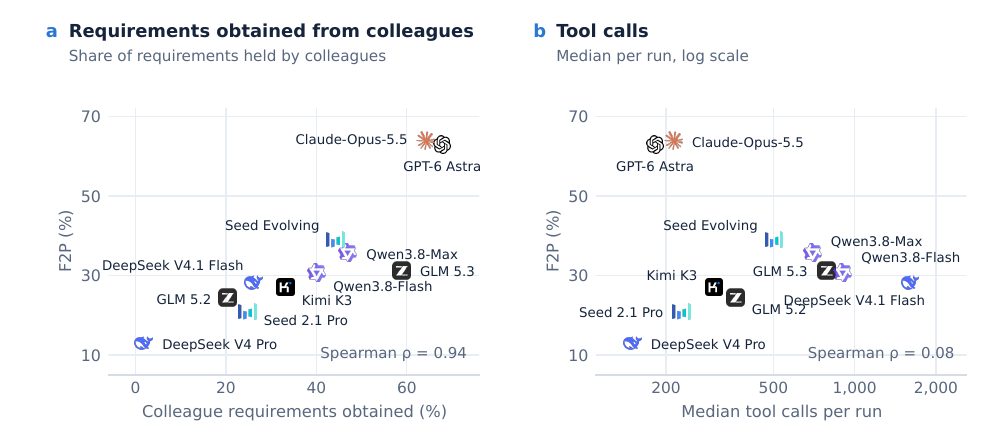}
\caption{\textbf{Requirement acquisition, tool calls, and delivery
across eleven models.}
(a)~F2P against the share of requirements held only by colleagues
that a colleague disclosed during the run.
(b)~F2P against the median number of tool calls per run, on a
logarithmic axis. Each model is shown by its developer's logo.
Spearman $\rho$ is computed over the eleven models.}
\label{fig:model-acquisition}
\end{figure}

\paragraph{Delivery tracks requirement acquisition.}
Figure~\ref{fig:model-acquisition} plots each model's F2P score
against the share of requirements it obtained from colleagues and
against its median number of tool calls per run.
Table~\ref{tab:model-acquisition} in \Cref{sec:appendix-acquisition}
lists all values, including colleague messages and wall-clock hours.
The colleague rate counts the requirements held only by colleagues
that a colleague disclosed during the run, as recorded by the
disclosure events of the colleague service.

Requirement acquisition is strongly associated with F2P
(Spearman $\rho = 0.94$), whereas tool-call counts show
little association ($\rho = 0.08$). GPT-6 Astra and
Claude-Opus-5.5 obtain 67.8\% and 64.2\% of colleague-held
requirements, respectively, and achieve the highest F2P
scores. In contrast, DeepSeek V4 Pro obtains only 1.8\%
and reaches 13.04\% F2P.

Greater interaction volume does not necessarily translate
into better requirement acquisition or delivery. Some models
exhaust the eight-hour execution budget with hundreds or
thousands of tool calls, yet obtain fewer requirements than
the leading models. These findings highlight the importance
of acquiring relevant organizational information, while
leaving open how effectively agents translate that
information into working implementations.

\paragraph{Interaction volume and context usage.}
Across nine models with available usage telemetry, additional analysis
of model calls, token usage, and context compactions shows no consistent
relationship between interaction volume and delivery. In particular,
large input-token totals may reflect repeated processing of retained
context rather than more extensive workplace investigation.
We provide the detailed usage comparisons in
Appendix~\ref{sec:appendix-acquisition}.

High acquisition also leaves an implementation gap. We examine it with
Oracle-Spec (Section~\ref{sec:seed-oracle}) and requirement-level
trajectories (Section~\ref{sec:opus-trajectories}).

\FloatBarrier
\begin{figure}[!t]
\centering
\includegraphics[width=\linewidth]{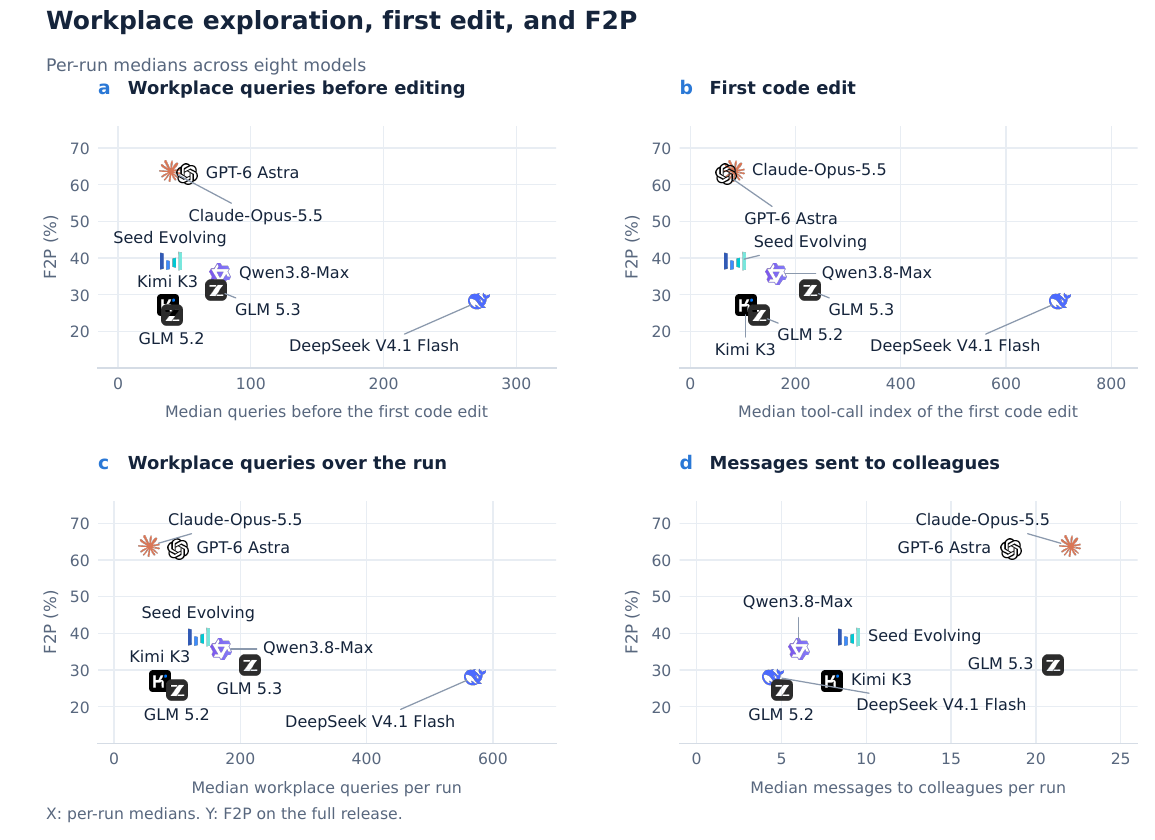}
\caption{\textbf{Workplace exploration, first code edit, and delivery.}
Full-release F2P against per-run medians of (a)~workplace queries
before the first code edit, (b)~the tool-call index of that edit,
(c)~workplace queries over the full run, and (d)~messages sent to
colleagues. All panels cover the same eight models.
All axes are linear. Each logo sits at the model's exact value; thin
leaders link names to logos in crowded regions.}
\label{fig:workplace-exploration}
\end{figure}

\paragraph{Workplace exploration and implementation.}
Figure~\ref{fig:workplace-exploration} examines how agents explore
organizational information and transition to code implementation.
Across eight models, workplace queries before the first edit and total
workplace queries show little rank association with F2P
(Spearman $\rho=\provisional{-0.10}$ and
$\rho=\provisional{-0.17}$, respectively). In contrast, a lower
tool-call index for the first code edit ($\rho=\provisional{-0.62}$)
and more frequent colleague communication ($\rho=\provisional{0.71}$)
are associated with higher F2P.

These patterns suggest that extensive investigation alone does not
ensure successful delivery. Leading models tend to begin implementation
after fewer tool calls while communicating more frequently with
colleagues. These cross-model associations do not establish that earlier
editing or more frequent messaging causes better outcomes. Together
with the requirement acquisition results, they highlight the importance
of integrating organizational information into the development process.

\FloatBarrier
\subsection{Effect of Supplying Complete Requirements}
\label{sec:seed-oracle}

For Seed~Evolving and GPT-6~Astra, supplying complete requirements
up front raises partial delivery while leaving complete task success
unchanged. The paired comparison holds the \provisional{101} tasks,
repositories, executable contracts, scaffold and execution budget fixed.

\begin{table}[!htbp]
\centering
\small
\setlength{\tabcolsep}{6pt}
\renewcommand{\arraystretch}{1.08}
\input{tables/table_seed_oracle}
\caption{\textbf{CoWork vs.\ Oracle-Spec for Seed~Evolving and GPT-6~Astra.}
Scores are percentages; gain is in percentage points.
Pooled metrics use the complete
\provisional{101}-task release contract.}
\label{tab:seed-oracle-comparison}
\end{table}

Providing the specification raises Seed~Evolving's F2P from
\provisional{39.09\%} to \provisional{55.76\%},
a gain of \provisional{16.67} percentage points
(Table~\ref{tab:seed-oracle-comparison}).
GPT-6~Astra's F2P rises by a similar amount, from
\provisional{62.91\%} to \provisional{78.84\%}, a gain of
\provisional{15.93} points. P2P moves in opposite directions:
Seed~Evolving's falls from \provisional{94.48\%} to
\provisional{89.50\%}, while GPT-6~Astra's rises from
\provisional{89.15\%} to \provisional{91.81\%}. With the
complete specification, Seed~Evolving implements more of the new
behavior and preserves less of the existing behavior.

The roughly \provisional{16}-point F2P gains reflect the combined
benefit of direct access to requirements and the effort saved in
searching for and confirming them. Complete task success nevertheless
remains unchanged: Seed~Evolving completes \provisional{0.99\%} of
tasks and GPT-6~Astra \provisional{7.92\%} in both conditions.
Even with every requirement in hand, Seed~Evolving leaves
\provisional{44\%} of F2P test nodes unsatisfied and GPT-6~Astra
\provisional{21\%}. These remaining failures occur during
implementation and verification, after requirements are available. The
Claude-Opus-5.5 analysis below follows individual requirements
through both stages.

\FloatBarrier
\subsection{Claude-Opus-5.5: Following Each Requirement to Delivery}
\label{sec:opus-trajectories}
\label{sec:qualitative}

We analyze \provisional{101} Claude-Opus-5.5 CoWork trajectories,
linking colleague interactions, requirement disclosures, executable
test outcomes, and final handoff reports. This analysis examines where
requirements remain unresolved as agents move from clarification to
implementation, verification, and delivery.

\subsubsection{Clarification: reaching owners and raising requirements}
\label{sec:loss}

\begin{figure}[!t]
\centering
\includegraphics[width=\linewidth]{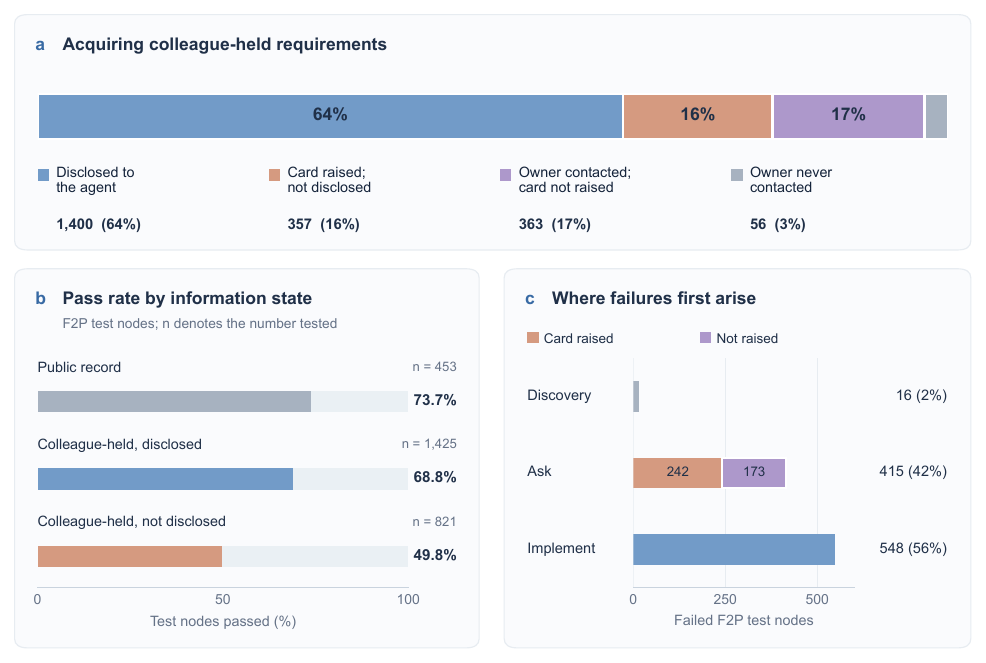}
\caption{\textbf{Requirement acquisition and loss in Claude-Opus-5.5.}
(a) Requirements held by colleagues, grouped by acquisition state.
(b) F2P test-node pass rates against the full release contract,
grouped by whether the corresponding requirement reached the agent.
(c) Earliest observed loss stage (discovery, ask, or implement) for failed
test nodes with linked requirements, evaluated against the full release contract.
The \emph{implement} category denotes unsatisfied behavior
after the requirement reached the agent.}
\label{fig:requirement-flow}
\end{figure}

\paragraph{Finding the right owner is not enough.}
Claude-Opus-5.5 directs its first question to the accountable owner
for a mean of \provisional{93\%} of queried cards across runs
(Table~\ref{tab:ask}), yet obtains only \provisional{64\%} of the
\provisional{2,176} colleague-held requirements
(Figure~\ref{fig:requirement-flow}). Of all colleague-held requirements,
\provisional{16\%} are raised without obtaining disclosure,
\provisional{17\%} remain unraised with owners already contacted about
other cards, and \provisional{3\%} belong to owners never contacted.
Thus, most missing requirements arise from incomplete clarification
with already-contacted owners rather than failures to identify the
right colleagues.

\paragraph{Acquired requirements remain difficult to deliver.}
F2P test nodes linked to disclosed colleague-held requirements pass at
\provisional{68.8\%}, compared with \provisional{49.8\%} for
undisclosed requirements. Within tasks containing both types, the mean
difference is \provisional{21.3} percentage points. However, among
failed F2P test nodes with linked requirements, \provisional{56\%}
are assigned to \emph{implement}: the requirement reached the agent,
but the tested behavior remains unsatisfied. Another \provisional{42\%}
are associated with requirements not successfully obtained through owner
clarification (\emph{ask}), and the remaining \provisional{2\%} are
attributed to \emph{discovery}. These results reveal gaps both in
obtaining requirements and in translating available information into
verified implementations.

\FloatBarrier
\subsubsection{Clarification over time: following up deferred questions}
\label{sec:follow-through}

\begin{figure}[H]
\centering
\includegraphics[width=\linewidth]{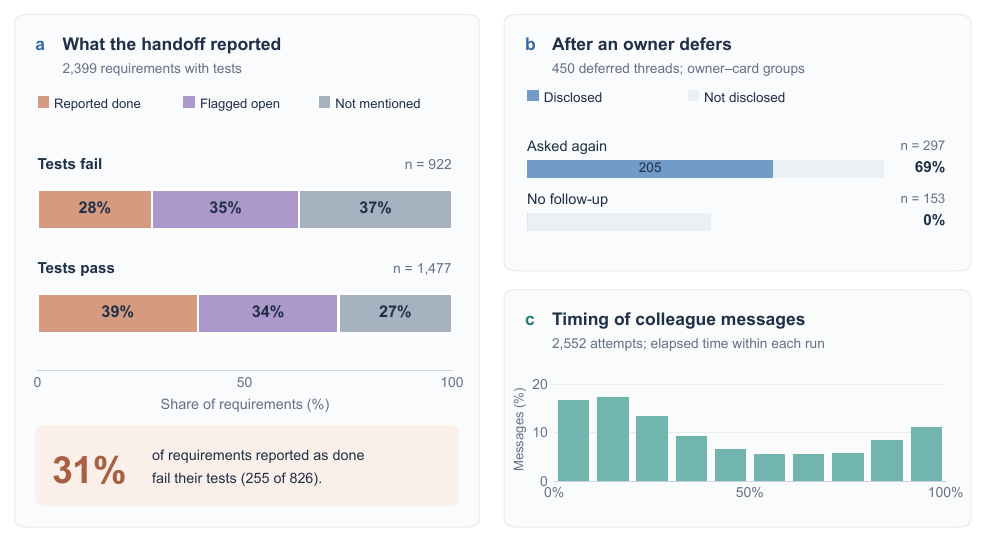}
\caption{\textbf{Follow-through and reporting in Claude-Opus-5.5.}
(a) Delivery claims compared with the test outcome of each requirement.
(b) Deferred threads grouped by owner and card, subsequent
follow-up and eventual disclosure.
(c) Timing of colleague messages within a run.
Message activity across the run can coexist with
individual threads receiving no follow-up.}
\label{fig:basics}
\end{figure}

\paragraph{Deferred questions remain unresolved despite continued activity.}
Among \provisional{450} owner--card threads in which a colleague
initially defers an answer, Claude-Opus-5.5 follows up on
\provisional{297}, eventually obtaining disclosure in \provisional{205}.
The remaining \provisional{153} threads receive no follow-up
(Figure~\ref{fig:basics}b). At the initial deferral, these abandoned
threads have a median of \provisional{7.1} hours of execution budget
remaining, and the agent continues working for a median of
\provisional{3.4} hours afterward.

Communication nevertheless continues late in execution: of
\provisional{2,552} colleague message attempts, \provisional{20\%}
occur during the final fifth of execution (Figure~\ref{fig:basics}c).
These aggregate patterns show that sustained communication can coexist
with individual questions left unresolved.

The PostCSS task (row 014) illustrates this difficulty. The agent
raises card CSS-227 and follows up with additional evidence, but
receives another deferral and eventually declares the streams complete
without resolving the question. All \provisional{16} failed F2P test
nodes in this run concern undisclosed colleague-held requirements.
The case illustrates how unresolved clarifications can persist through
delivery despite earlier attempts to obtain an answer.

\subsubsection{Implementation and verification of obtained requirements}
\label{sec:implementation-verification}

\paragraph{Requirement acquisition does not guarantee implementation.}
The carapace-bin task (row 001) illustrates this gap.
Claude-Opus-5.5 obtains all \provisional{25} colleague-held requirements,
yet fails \provisional{22} of \provisional{35} F2P test nodes.
The trajectory reviewer identifies incorrect usages, flags, and
candidate descriptions, along with an unchecked build failure.
Despite acquiring the requirements, the agent leaves substantial parts
of the executable contract unsatisfied.

\begin{figure}[H]
\centering
\includegraphics[width=\linewidth]{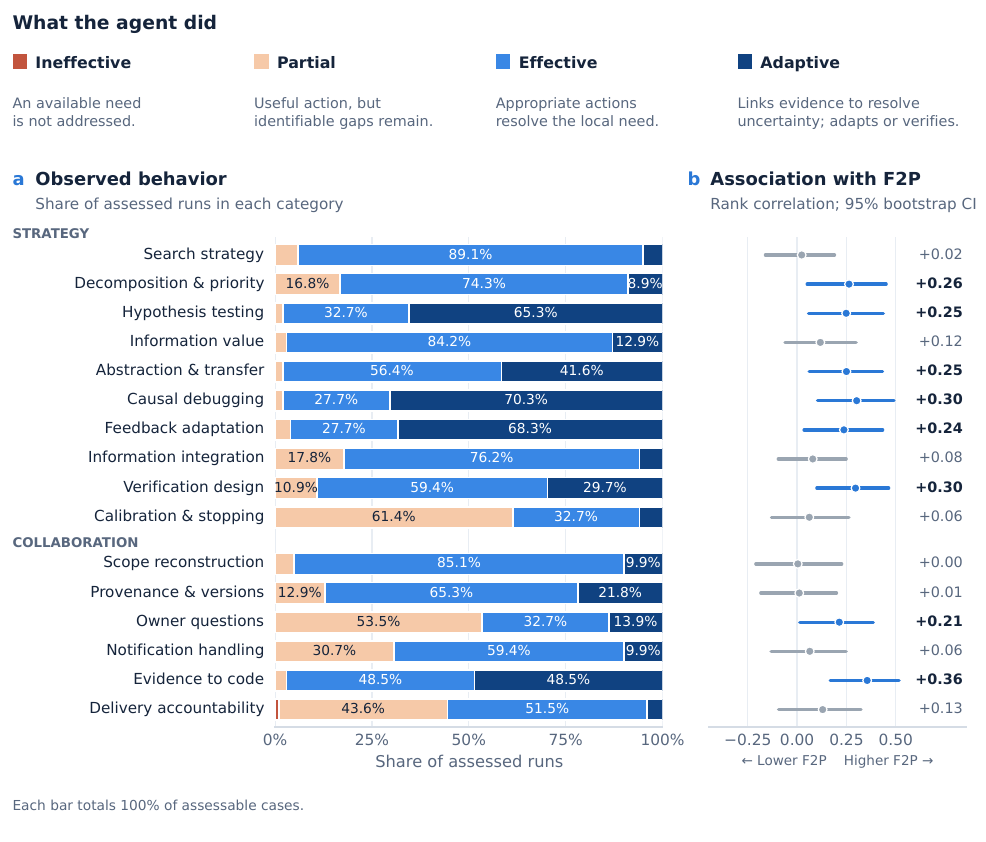}
\caption{\textbf{Claude-Opus-5.5 behavior profile.}
(a) Percentages of assessable runs by observed behavior.
\emph{Ineffective} actions fail to address an available need;
\emph{partial} handling leaves identifiable gaps;
\emph{effective} handling resolves the local need;
and \emph{adaptive} handling connects evidence across
steps to resolve uncertainty and adapt or verify.
Each bar is normalized over assessable cases for
that dimension and evaluates behavior in its local interaction context.
(b) Exploratory Spearman associations with per-task F2P,
with 95\% bootstrap intervals.}
\label{fig:behaviour}
\end{figure}

\paragraph{Translating evidence into working code.}
Figure~\ref{fig:behaviour} examines exploratory associations between
the trajectory reviewer's behavioral assessments and per-task F2P.
Among the assessed dimensions, evidence-to-code translation shows the
strongest association with F2P (Spearman $\rho=\provisional{+0.36}$),
followed by causal debugging and verification design
(both $\rho=\provisional{+0.30}$). The correlations for
colleague-message count and tool-call count are
$\rho=\provisional{+0.21}$ and $\rho=\provisional{-0.21}$, respectively.

These exploratory results suggest that effective use of acquired
information matters beyond the amount of interaction. Agents may
obtain the necessary requirements yet fail to incorporate them
correctly into code or verify the resulting behavior.

\subsubsection{Handoff: stopping decisions and reported status}
\label{sec:basics}

\paragraph{Stopping before the work is complete.}
The trajectory reviewer identifies \provisional{49} premature stops
in which unresolved in-scope requirements still have actionable next
steps (Figure~\ref{fig:effort-score}e in
Appendix~\ref{sec:appendix-run-constraints}). Among the
\provisional{48} autonomous stops in this group, the median remaining
execution budget is \provisional{2.8} hours. These observations suggest
that the agent may terminate assignments despite having both
outstanding work and opportunities to continue.

\paragraph{Completion claims diverge from test outcomes.}
Among \provisional{826} requirements reported as completed at handoff,
\provisional{255} (\provisional{31\%}) fail their corresponding tests
(Figure~\ref{fig:basics}a). Conversely, many unsatisfied requirements
are omitted from the final report rather than explicitly identified
as outstanding.

The HotswapAgent task (row 021) illustrates this discrepancy: the agent
declares the assignment complete while its own notes still list
\provisional{8} pending requirements. Together, these observations
reveal difficulties in determining when an assignment is complete
and communicating its remaining obligations accurately.

\FloatBarrier
\subsection{Tracking Requirements Through an Assignment}
\label{sec:expert-workflow}

Across the model comparisons, Oracle-Spec experiment, and trajectory
analysis, requirements remain unsatisfied at multiple stages of an
assignment. Supplying complete requirements improves partial delivery
without ensuring whole-task success, while the trajectory analysis
exposes gaps in implementation, verification, and handoff. Together,
these findings suggest that reliable delivery requires maintaining
the state of individual requirements throughout the engineering workflow.

One direction is to explicitly track each requirement's accountable
owner, unresolved questions, implementation status, verification
evidence, and handoff claims. Such tracking could help agents revisit
outstanding dependencies and distinguish implemented changes from
verified outcomes. Whether this approach improves end-to-end delivery
remains an open question.

%% file: tables/table_partial_models.tex
\begin{tabular}{@{}lrrr@{}}
\toprule
Model & F2P (\%) & P2P (\%) & Pass rate (\%) \\
\midrule
\iconclaude Claude-Opus-5.5 & \rankfirst{\provisional{63.84}} & \provisional{93.24} & \ranksecond{\provisional{6.93}} \\
\addlinespace[3pt]
\icongpt GPT-6 Astra & \ranksecond{\provisional{62.91}} & \provisional{89.15} & \rankfirst{\provisional{7.92}} \\
\addlinespace[3pt]
\iconseed Seed Evolving & \rankthird{\provisional{39.09}} & \ranksecond{\provisional{94.48}} & \rankthird{\provisional{0.99}} \\
\addlinespace[3pt]
\iconqwen Qwen3.8-Max & \provisional{35.70} & \provisional{91.80} & \provisional{0.00} \\
\addlinespace[3pt]
\iconglm GLM 5.3 & \provisional{31.27} & \provisional{87.72} & \rankthird{\provisional{0.99}} \\
\addlinespace[3pt]
\iconqwen Qwen3.8-Flash & \provisional{30.70} & \provisional{93.40} & \provisional{0.00} \\
\addlinespace[3pt]
\icondeepseek DeepSeek V4.1 Flash & \provisional{28.20} & \rankthird{\provisional{94.30}} & \provisional{0.00} \\
\addlinespace[3pt]
\iconkimi Kimi K3 & \provisional{27.08} & \provisional{84.88} & \rankthird{\provisional{0.99}} \\
\addlinespace[3pt]
\iconglm GLM 5.2 & \provisional{24.40} & \rankfirst{\provisional{95.20}} & \provisional{0.00} \\
\addlinespace[3pt]
\iconseed Seed 2.1 Pro & \provisional{21.01} & \provisional{81.85} & \provisional{0.00} \\
\addlinespace[3pt]
\icondeepseek DeepSeek V4 Pro & \provisional{13.04} & \provisional{90.21} & \provisional{0.00} \\
\bottomrule
\end{tabular}

%% file: tables/table_seed_oracle.tex
\begin{tabular}{@{}lrrrrrr@{}}
\toprule
& \multicolumn{3}{c}{Seed~Evolving} & \multicolumn{3}{c}{GPT-6~Astra} \\
\cmidrule(lr){2-4}\cmidrule(l){5-7}
Metric & CoWork & Oracle & Gain & CoWork & Oracle & Gain \\
\midrule
F2P & \provisional{39.09} & \provisional{55.76} & \provisional{+16.67} & \provisional{62.91} & \provisional{78.84} & \provisional{+15.93} \\
\addlinespace[2pt]
P2P & \provisional{94.48} & \provisional{89.50} & \provisional{$-$4.98} & \provisional{89.15} & \provisional{91.81} & \provisional{+2.66} \\
\addlinespace[2pt]
Pass rate & \provisional{0.99} & \provisional{0.99} & \provisional{+0.00} & \provisional{7.92} & \provisional{7.92} & \provisional{+0.00} \\
\bottomrule
\end{tabular}

%% file: content/7_Limitations.tex
\section{Conclusion}

We introduced \bench, a benchmark that evaluates coding agents
working within software organizations, where requirements must be
recovered from colleagues and workplace records before being
implemented and verified. Our evaluation reveals a substantial gap
between partial implementation and complete task delivery. Agents
frequently leave clarifications unresolved and, after obtaining
requirements, may still fail to satisfy executable tests or report
completion inconsistently with verified outcomes.

These findings highlight sustained requirement tracking as a key
challenge for autonomous software engineering. Future agents will
need to maintain continuity across collaboration, implementation,
verification, and handoff to support reliable delegation of
engineering work. \bench provides a setting for studying these
capabilities and can be extended to evaluate teams of autonomous
programming agents.

%% file: content/D_Contributions.tex

\section{Contributions}
\label{sec:appendix-contributions}

\paragraph{Leading Authors}
Xinping Lei

\paragraph{Core Contributors}
Minghao Liu, Ruijie Zhu, Xiaohui Hu

\paragraph{Contributors}
Xingqiao Li, Ao Zhang, Wenhao Chai, Han Li, Tao Wang, Jingyuan Zhang, Enduo Zhao, Yunfeng Shi, Yiming Liang, Zetao Yang, Xingjian Wang, Yifan Yao, Xinkai Ma, Yingru Li, Tianhan Yang, Huan Zhou, Weiran Shi, Shen Yan

\paragraph{Corresponding Authors}
 Jiaheng Liu, Stephen Huang, Eli Cheung, Weihao Xuan

%% file: content/A_Appendix.tex

\section{Broader Benchmark Lineage}
\label{sec:appendix-related}

Table~\ref{tab:related} compares repository and interactive benchmarks
by their research focus, information sources and evaluation design.
Table~\ref{tab:recent} provides a more detailed comparison of recent
benchmarks in terms of task goals, initial inputs and information processes.
SWE-bench \citep{jimenez2024swebench} established patch evaluation in real repositories
using F2P/P2P tests. SWE-bench Verified
\citep{swebenchverified}, SWE-bench Multimodal
\citep{yang2024swebenchmm}, Multi-SWE-bench
\citep{zan2025multiswebench} and SWE-bench Multilingual
\citep{yang2025swesmith} extend validation quality, modalities
and language coverage. SWE-bench Pro \citep{swebenchpro2025}
evaluates larger changes, SWE-Lancer \citep{miserendino2025swelancer}
draws on paid freelance assignments, and Active-SWE
\citep{activeswe2026} evaluates proactive bug discovery
from repository state.

\begin{table}[!htbp]
\centering
\small
\setlength{\tabcolsep}{4pt}
\input{tables/table_related}
\caption{Comparison of repository and interactive benchmarks by
research focus, information sources and evaluation design.}
\label{tab:related}
\end{table}

\FloatBarrier
\Needspace{0.75\textheight}
\section{Released Repositories}
\label{sec:appendix-repositories}

Table~\ref{tab:domains} lists the \nrepos upstream repositories
represented in the \ntasks released tasks, grouped by application domain.

\begin{table}[!htbp]
\centering
\small
\input{tables/table_domains}
\caption{The \ntasks released tasks by application domain and
upstream repository.}
\label{tab:domains}
\end{table}



\section{Evidence Case Examples}
\label{sec:appendix-cases}

\begin{figure}[!htbp]
\centering
\includegraphics[width=\linewidth]{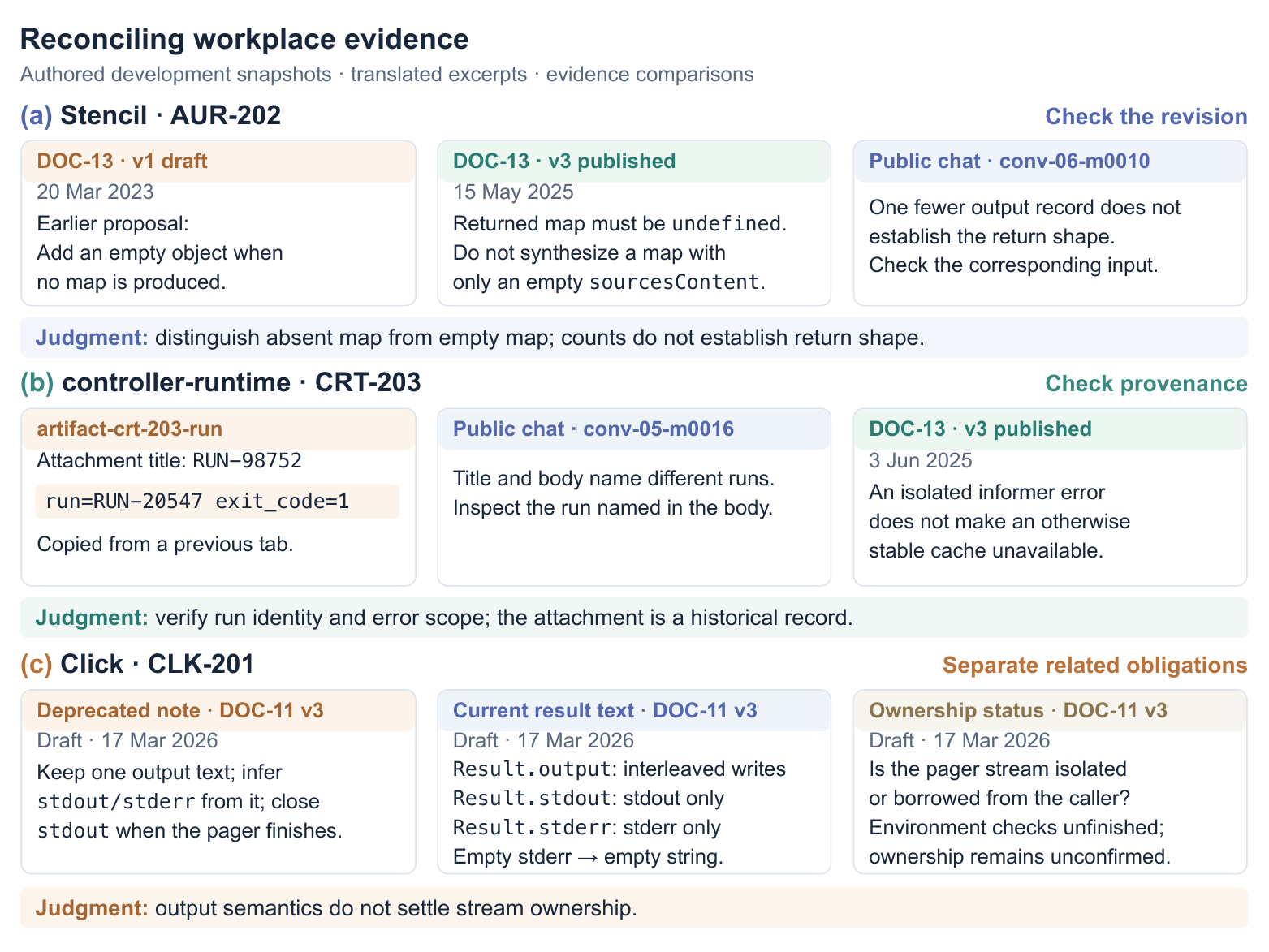}
\caption{Examples of evidence assessment in workplace records, with
source card, document, version and message identifiers.
(a) A published correction supersedes an earlier return shape.
(b) Run identifiers expose a discrepancy in copied terminal output.
(c) Output semantics and an unresolved question about stream ownership.}
\label{fig:case-examples}
\end{figure}

Figure~\ref{fig:case-examples} illustrates the judgments about document
history, evidence provenance and scope discussed in
Section~\ref{sec:obstacles}. The examples use workplace records
from Stencil, controller-runtime and Click. The Click records also
appear in the opening comparison in Figure~\ref{fig:case-contrast},
where the issue identifies the output capture requirement in
\texttt{DOC-11}.

\paragraph{Stencil: preserving an undefined map field.}
Card \texttt{AUR-202} leads to \texttt{DOC-13}, whose published
correction supersedes an earlier proposal to fill in an empty map.
When \texttt{map} is \texttt{undefined}, the chunk must still
retain its other output information. The linked run excerpt
reports 22 versus 21 output records but provides no record shapes.
The public reply requests the corresponding input to determine
whether the relevant map was absent or empty.

\paragraph{controller-runtime: verifying run identity and error scope.}
Around \texttt{CRT-203}, a public message identifies a mismatch
between the attachment title and the run identified in copied
terminal output. The linked \texttt{DOC-13} also corrects an earlier
interpretation that an informer error affected the entire cache.
The run identifiers establish which execution produced the
observation. Assessing the error's scope requires separate evidence
about the state of the otherwise stable cache.

\paragraph{Click: checking output semantics and stream ownership.}
The handoff on \texttt{CLK-201} points to \texttt{DOC-11}.
Its third version, marked draft, distinguishes interleaved output
from output attributed to individual sources and retains an earlier,
deprecated simplification. An incomplete environment check leaves
the pager's stream ownership unresolved. Resolving this question
requires identifying the component that supplied the stream and
whether the stream remains in use after return.

\FloatBarrier
\Needspace{0.62\textheight}
\section{Per-Task Statistics}
\label{sec:appendix-stats}

Table~\ref{tab:stats} reports statistics on software changes,
workplace records and information design across the released tasks.

\begin{table}[!htbp]
\centering
\small
\input{tables/table_stats}
\caption{Statistics across the \ntasks released tasks.
Production lines and files count source code, excluding tests,
documentation, configuration, generated files, fixtures,
lockfiles and vendored code.}
\label{tab:stats-full}
\label{tab:stats}
\end{table}

\section{Supplementary Claude-Opus-5.5 Diagnostics}
\label{sec:appendix-results}

The following tables provide additional results from the
Claude-Opus-5.5 trajectory analysis in
Section~\ref{sec:opus-trajectories}. They report information
acquisition behavior and performance across task strata.

\begin{table}[!htbp]
\centering
\small
\setlength{\tabcolsep}{4pt}
\input{tables/table_observed_ask}
\caption{\textbf{Claude-Opus-5.5 acquisition behavior.}
The first row reports means across runs; the second reports
Spearman correlations between each acquisition measure and
task F2P. Ask rate, targeting, gain per question, follow-through,
redundancy and unread disclosures are defined in
Section~\ref{sec:experiments}. All means are percentages
except gain per question.}
\label{tab:ask}
\end{table}

\begin{table}[!htbp]
\centering
\small
\input{tables/table_observed_strata}
\caption{\textbf{Stratified Claude-Opus-5.5 diagnostics.}
F2P is the mean of per-task F2P fractions within each stratum.
Table~\ref{tab:partial-models} reports a pooled F2P score.
Passed tasks counts tasks satisfying both F2P and P2P.
Disclosed is the mean proportion of requirements held by
colleagues that the agent obtained. Edit size counts production
lines added by the reference change.}
\label{tab:strata}
\end{table}

\FloatBarrier
\subsection{Run Activity, Stopping, and Assignment Constraints}
\label{sec:appendix-run-constraints}

\begin{figure}[!htbp]
\centering
\includegraphics[width=\linewidth]{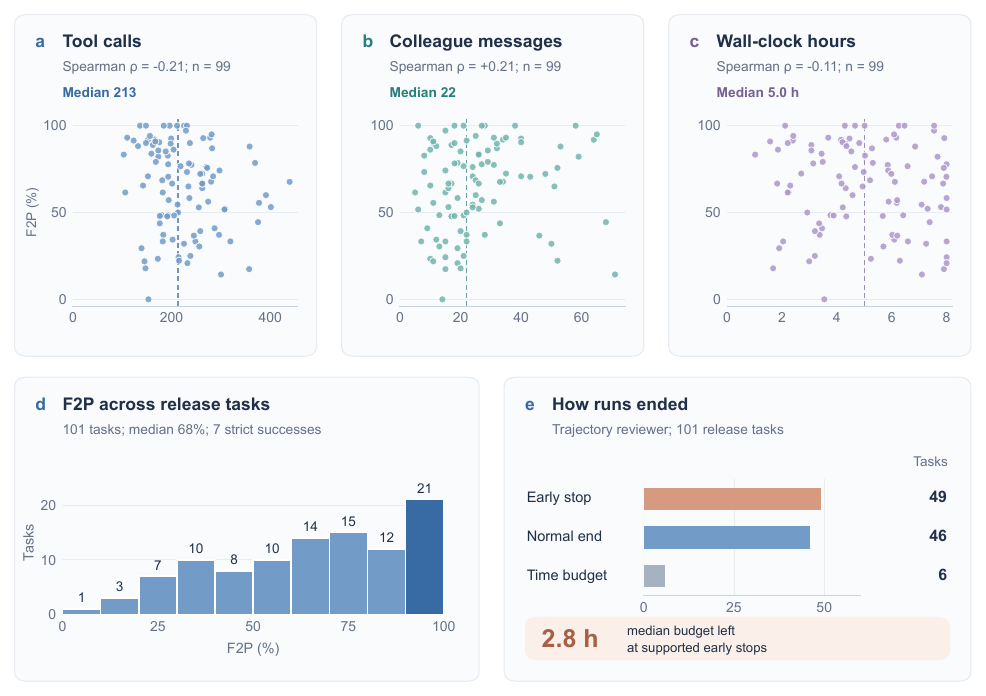}
\caption{\textbf{Effort, outcomes and stopping in Claude-Opus-5.5.}
(a--c) Tool calls, colleague messages and wall-clock hours
against F2P fractions with per-run telemetry.
(d) Distribution of F2P fractions over the
101-task release cohort.
(e) Judgments by the trajectory reviewer of how runs ended,
including stops with known, actionable gaps.}
\label{fig:effort-score}
\end{figure}

\paragraph{Run activity.}
Figure~\ref{fig:effort-score} summarizes per-run tool calls, colleague
messages, wall-clock usage, F2P outcomes, and stopping judgments by
the trajectory reviewer.
A median Claude-Opus-5.5 run contacts \provisional{6} colleagues
and uses \provisional{5.0} hours of the \provisional{8}-hour
execution budget. Elapsed time shows little rank association with
per-task F2P (Figure~\ref{fig:effort-score}c).

\paragraph{Changes to existing tests.}
Among \provisional{94} assignment briefs that prohibit changes to
existing tests, \provisional{15} submitted patches change or remove
lines in existing test sources, violating the assignment constraints.

\FloatBarrier
\section{Run Behavior and Requirement Acquisition}
\label{sec:appendix-acquisition}

Table~\ref{tab:model-acquisition} lists the values plotted in
Figure~\ref{fig:model-acquisition}, together with the median
number of colleague messages and wall-clock hours per run.
The run median computes the colleague rate for each run and takes
the median, so it can rank models differently from the pooled rate.

\begin{table}[!htbp]
\centering
\small
\setlength{\tabcolsep}{4.5pt}
\input{tables/table_model_acquisition}
\caption{\textbf{Run behavior and requirement acquisition.}
Messages are sent to colleagues through \texttt{chat\_send};
tool calls include terminal commands and workspace tools;
hours are agent wall-clock time, capped at \provisional{8}.
Colleague: share of requirements held only by colleagues that
were disclosed. All: disclosed colleague requirements plus public
requirements. Run median: median of the per-run colleague rate.
Rows are ordered by the colleague rate.}
\label{tab:model-acquisition}
\end{table}

\FloatBarrier
\subsection{Interaction Volume and Context Usage}
\label{sec:appendix-usage}

\begin{figure}[!htbp]
\centering
\includegraphics[width=\linewidth]{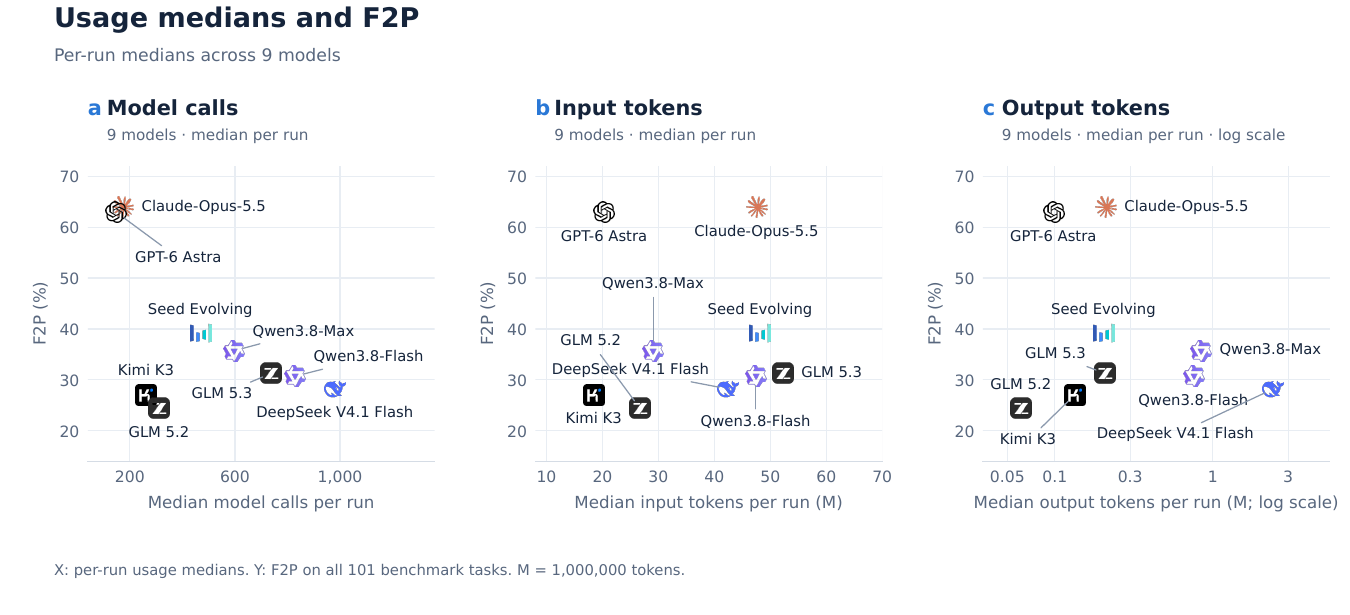}
\caption{\textbf{Model calls, token usage, and delivery.}
F2P on the full release against per-run medians of (a)~model calls,
(b)~input tokens, and (c)~output tokens.
Each point is a model; token axes use millions (M), with a logarithmic
scale in (c). Each logo sits at the model's exact value; thin
leaders link names to logos in crowded regions.}
\label{fig:usage-medians}
\end{figure}

\begin{figure}[!htbp]
\centering
\includegraphics[width=\linewidth]{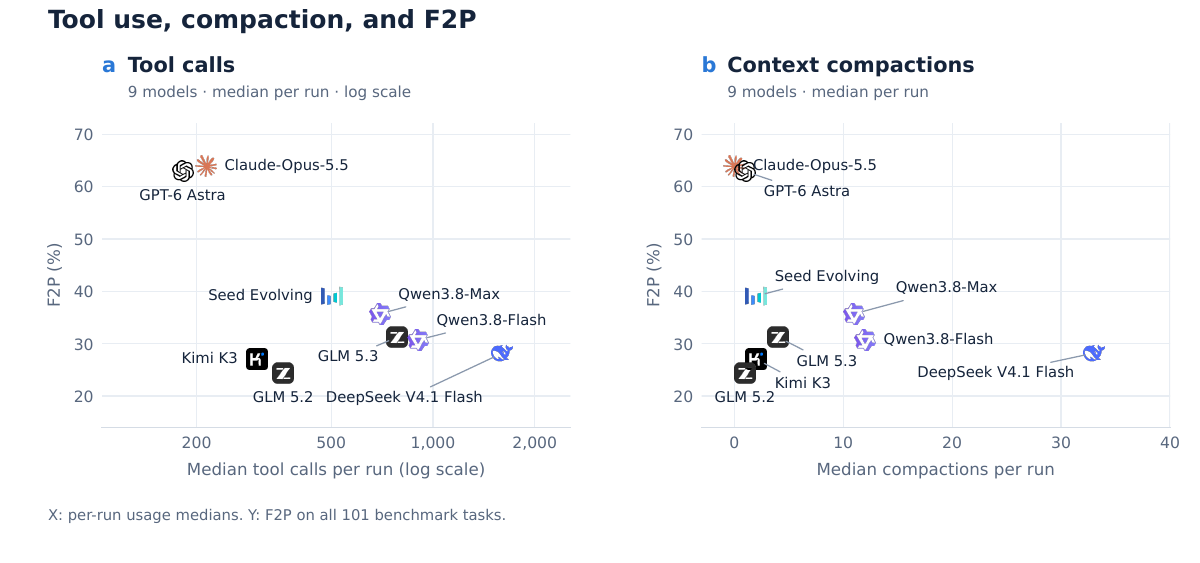}
\caption{\textbf{Tool use, context compaction, and delivery.}
Full-release F2P against per-run medians of (a)~tool calls and
(b)~context compactions for nine models. Tool calls use a logarithmic
axis; compactions use a linear axis to include zero.
Each logo sits at the model's exact value; thin leaders link names to
logos in crowded regions.}
\label{fig:usage-tools}
\end{figure}

Figures~\ref{fig:usage-medians} and~\ref{fig:usage-tools} compare
per-run medians of model calls, token usage, tool calls, and context
compactions with full-release F2P across the nine models with available
usage telemetry. Token-usage medians are unavailable for
DeepSeek~V4~Pro and Seed~2.1~Pro.

The models exhibit different interaction patterns. Some lower-scoring
models terminate after relatively few calls, while others sustain
extensive tool use without achieving correspondingly higher F2P scores.
Claude-Opus-5.5 and GPT-6~Astra achieve the highest F2P scores despite
making relatively few model and tool calls.

Input-token totals also depend on context management.
Claude-Opus-5.5 processes a median of \provisional{47.57} million
input tokens without context compaction, compared with
\provisional{20.32} million for GPT-6~Astra. These totals include
repeated processing of retained context and therefore do not directly
measure the extent of workplace exploration.

%% file: tables/table_related.tex
\begin{tabular}{@{}p{0.18\linewidth}p{0.23\linewidth}p{0.25\linewidth}p{0.26\linewidth}@{}}
\toprule
Benchmark & Focus & Information setting & Evaluation / control \\
\midrule
SWE-bench \citep{jimenez2024swebench}
  & Issue resolution & Issue and repository & F2P/P2P tests \\[4pt]
SWE-bench Pro \citep{swebenchpro2025}
  & Complex software changes & Issue and repository & Repository tests \\[4pt]
SWE-Lancer \citep{miserendino2025swelancer}
  & Paid freelance work & Task description & End-to-end tests for implementation tasks \\[4pt]
Active-SWE \citep{activeswe2026}
  & Proactive bug discovery and fixing & Repository without issue reports
  & Bug discovery and repair evaluation \\[4pt]
TheAgentCompany \citep{xu2024agentcompany}
  & Professional workflows, including SWE & Company services and multiple colleagues
  & Programmatic checkpoints; some model-based criteria \\[4pt]
LHAW \citep{pu2026lhaw}
  & Controlled underspecification & Removed goals, constraints, inputs or context
  & Inherited evaluators and agent trials \\[4pt]
Ask or Assume \citep{askorassume2026}
  & Clarification under uncertainty & Full issue held by a user simulator
  & Full, hidden and interactive conditions \\[4pt]
Dialogue SWE-Bench \citep{dialogueswe2026}
  & Coding through dialogue & Persona-grounded user simulator
  & Task outcomes and dialogue quality \\[4pt]
SWE-RPG \citep{swerpg2026}
  & Requirement and planning diagnosis & Issue and repository
  & Executable patches and intermediate reference annotations \\[4pt]
ICAE-Bench \citep{icaebench2026}
  & Interactive project building & Fuzzy requirements and a user simulator
  & Behavioral tests and multiple diagnostics \\[4pt]
$\tau^{\tau}$-Bench \citep{shi2026hypertau}
  & Agent construction & Business artifacts and a client
  & Held-out user simulations; serving constraints \\
\midrule
\bench (ours) & Autonomous task completion
  & Distributed evidence, multiple owners and late notices
  & \textbf{Same-task oracle control}, F2P/P2P tests and recovery traces \\
\bottomrule
\end{tabular}

%% file: tables/table_domains.tex
\begin{tabular}{@{}lp{0.035\linewidth}p{0.66\linewidth}@{}}
\toprule
Domain & $n$ & Upstream repositories \\
\midrule
Web \& Frontend & 26 & ada-url/ada, api-platform/core, collective/icalendar, emberjs/ember.js, feedbin/feedbin, fomantic/fomantic-ui, gioxx/marvelloussuspender, googleapis/google-api-php-client, hyperf/hyperf, ing-bank/lion, ionic-team/stencil, jasontaylordev/cleanarchitecture, jeremydaly/lambda-api, jnunemaker/httparty, jquery/jquery, kimai/kimai, mika-cn/maoxian-web-clipper, nolimits4web/swiper, python/pythondotorg, rubenlagus/telegrambots, saloonphp/saloon, shane32/qrcoder, tanstack/query, urllib3/urllib3, vacanza/holidays, webrecorder/pywb \\
\addlinespace[2pt]
Dev Tools \& Compilers & 22 & ansible/ansible-lint, carapace-sh/carapace-bin, cargo-generate/cargo-generate, cocoapods/xcodeproj, embarkstudios/cargo-deny, firebase/firebase-tools, git-lfs/git-lfs, hotswapprojects/hotswapagent, isomorphic-git/isomorphic-git, jctools/jctools, jest-community/jest-extended, junit-team/junit-examples, kpdecker/jsdiff, mdx-js/mdx, microsoft/vscode-js-debug, nedbat/coveragepy, pallets/click, postcss/postcss, psf/black, serde-rs/serde, stoplightio/elements, thought-machine/please \\
\addlinespace[2pt]
Data \& Databases & 11 & alibaba/jetcache, clockworklabs/spacetimedb, dolthub/go-mysql-server, elastic/elasticsearch-rails, friendsofsymfony/foselasticabundle, kotlin-orm/ktorm, mongodb/laravel-mongodb, qishibo/anotherredisdesktopmanager, rubyzip/rubyzip, ruflin/elastica, webtorrent/webtorrent \\
\addlinespace[2pt]
Games \& Graphics & 10 & caffeinemc/sodium, hubs-foundation/hubs, mezz/justenoughitems, mikke89/rmlui, multitheftauto/mtasa-blue, nasa-ammos/3dtilesrendererjs, sergio0694/computesharp, steamre/steamkit, tencent/puerts, underminersteam/undertalemodtool \\
\addlinespace[2pt]
Desktop Applications & 7 & betterdiscord/betterdiscord, bobotig/python-mss, electronnet/electron.net, ipfs/ipfs-desktop, openutau/openutau, pimutils/khal, sozi-projects/sozi \\
\addlinespace[2pt]
Cloud \& Infrastructure & 5 & boto/boto3, brightercommand/brighter, kubermatic/kubeone, robusta-dev/robusta, stackstorm/st2 \\
\addlinespace[2pt]
Finance \& Payments & 5 & btcsuite/btcd, ledgerhq/ledger-live, mempool/mempool, stripe/stripe-php, stripe/stripe-ruby \\
\addlinespace[2pt]
ML \& AI Systems & 4 & huggingface/transformers.js, infiniflow/ragflow, lmcache/lmcache, onnx/onnx \\
\addlinespace[2pt]
Security \& Identity & 4 & apache/caldera, bubka/2fauth, crytic/slither, pglombardo/passwordpusher \\
\addlinespace[2pt]
Mobile Applications & 3 & luoyesiqiu/dpt-shell, mygod/vpnhotspot, react-native-async-storage/async-storage \\
\addlinespace[2pt]
Observability & 2 & errbit/errbit, newrelic/newrelic-ruby-agent \\
\addlinespace[2pt]
Scientific Computing & 2 & angularsen/unitsnet, google-deepmind/mujoco \\
\bottomrule
\end{tabular}

%% file: tables/table_stats.tex
\begin{tabular}{lrrrrr}
\toprule
Quantity per task & Mean & Median & P25 & P75 & Max \\
\midrule
\multicolumn{6}{l}{\textit{Code task}} \\
\quad Verifier test nodes & 32.3 & 32 & 28 & 35 & 68 \\
\quad Fail-to-pass test nodes & 26.7 & 27 & 23 & 31 & 67 \\
\quad Pass-to-pass test nodes & 5.6 & 4 & 3 & 8 & 19 \\
\quad Production files in golden patch & 97.5 & 60 & 37 & 133 & 524 \\
\quad Production lines added & 4,590.3 & 2,955 & 1,687 & 5,645 & 26,605 \\
\quad Production lines deleted & 1,752.0 & 1,069 & 468 & 2,217 & 13,232 \\
\addlinespace
\multicolumn{6}{l}{\textit{Workplace}} \\
\quad Colleagues & 26.0 & 26 & 26 & 26 & 26 \\
\quad Board cards & 102.3 & 102 & 100 & 104 & 107 \\
\quad Card comments & 62.2 & 63 & 58 & 66 & 80 \\
\quad Documents & 50.5 & 50 & 48 & 52 & 56 \\
\quad Chat messages & 404.3 & 405 & 399 & 410 & 417 \\
\quad Merge requests & 11.8 & 12 & 11 & 13 & 14 \\
\quad MR review notes & 27.8 & 28 & 26 & 30 & 35 \\
\quad Run artifacts & 49.9 & 50 & 47 & 53 & 64 \\
\addlinespace
\multicolumn{6}{l}{\textit{Information design}} \\
\quad Scored behavioral facts & 26.9 & 28 & 24 & 30 & 32 \\
\quad Contract atoms pinned by tests & 28.6 & 28 & 18 & 34 & 88 \\
\quad Static contract atoms & 11.1 & 5 & 5 & 14 & 58 \\
\quad Facts held by an owner & 22.1 & 23 & 19 & 25 & 28 \\
\quad Progress-gated deliveries & 8.0 & 8 & 8 & 8 & 8 \\
\bottomrule
\end{tabular}

%% file: tables/table_observed_ask.tex
\begin{tabular}{@{}lrrrrrr@{}}
\toprule
& Ask rate & Targeting & Gain/Q & Follow-thr. & Redundancy & Unread \\
\midrule
Claude-Opus-5.5 (\textsc{CoWork}) & 80.4 & 93.3 & 0.64 & 78.3 & 13.7 & 0.2 \\
Spearman $\rho$ with F2P & +0.11 & +0.14 & +0.04 & +0.13 & +0.14 & +0.18 \\
\bottomrule
\end{tabular}

%% file: tables/table_observed_strata.tex
\begin{tabular}{@{}llrrrr@{}}
\toprule
Axis & Stratum & $n$ & F2P (\%) & Passed tasks & Disclosed (\%) \\
\midrule
 \multirow{3}{*}{Verifier test nodes} & $< 28$ & 21 & 70.1 & 3 & 76.5 \\
  & $28$--$35$ & 55 & 62.5 & 2 & 61.3 \\
  & $> 35$ & 25 & 64.7 & 2 & 61.9 \\
\addlinespace[2pt]
 \multirow{3}{*}{Edit size} & $<1$k production lines & 6 & 67.6 & 0 & 75.0 \\
  & $1$k--$5$k & 58 & 66.0 & 5 & 65.4 \\
  & $>5$k & 37 & 62.0 & 2 & 61.7 \\
\addlinespace[2pt]
 \multirow{5}{*}{Language} & Python & 17 & 69.3 & 1 & 62.2 \\
  & TypeScript / JavaScript & 32 & 57.3 & 1 & 64.1 \\
  & PHP / Ruby & 19 & 68.4 & 1 & 68.2 \\
  & Go / Rust & 11 & 56.5 & 0 & 72.6 \\
  & C++ / C\# / Java / Kotlin & 22 & 72.3 & 4 & 60.2 \\
\addlinespace[2pt]
 \multirow{5}{*}{Domain} & Web \& frontend & 26 & 75.8 & 4 & 72.2 \\
  & Dev tools \& compilers & 22 & 63.9 & 2 & 61.8 \\
  & Data \& databases & 11 & 70.2 & 1 & 67.6 \\
  & Games \& graphics & 10 & 48.9 & 0 & 58.5 \\
  & Other domains & 32 & 58.9 & 0 & 61.4 \\
\bottomrule
\end{tabular}

%% file: tables/table_model_acquisition.tex
\begin{tabular}{@{}lrrrrrrr@{}}
\toprule
& \multicolumn{3}{c}{Median per run} & \multicolumn{3}{c}{Requirements obtained (\%)} & \\
\cmidrule(lr){2-4}\cmidrule(lr){5-7}
Model & Messages & Tool calls & Hours & Colleague & All & Run median & F2P (\%) \\
\midrule
\icongpt GPT-6 Astra & \provisional{18.5} & \provisional{182} & \provisional{3.00} & \provisional{67.8} & \provisional{75.3} & \provisional{71.6} & \provisional{62.91} \\
\iconclaude Claude-Opus-5.5 & \provisional{22} & \provisional{213} & \provisional{5.01} & \provisional{64.2} & \provisional{70.7} & \provisional{72.0} & \provisional{63.84} \\
\iconglm GLM 5.3 & \provisional{21} & \provisional{785} & \provisional{2.46} & \provisional{58.8} & \provisional{66.3} & \provisional{60.9} & \provisional{31.27} \\
\iconqwen Qwen3.8-Max & \provisional{6} & \provisional{696} & \provisional{8.00} & \provisional{46.9} & \provisional{56.6} & \provisional{48.7} & \provisional{35.70} \\
\iconseed Seed Evolving & \provisional{9} & \provisional{503} & \provisional{4.46} & \provisional{44.3} & \provisional{50.6} & \provisional{38.1} & \provisional{39.09} \\
\iconqwen Qwen3.8-Flash & \provisional{10} & \provisional{902} & \provisional{5.53} & \provisional{40.1} & \provisional{51.0} & \provisional{38.1} & \provisional{30.70} \\
\iconkimi Kimi K3 & \provisional{8} & \provisional{301} & \provisional{1.41} & \provisional{33.1} & \provisional{45.2} & \provisional{27.3} & \provisional{27.08} \\
\icondeepseek DeepSeek V4.1 Flash & \provisional{4.5} & \provisional{1603} & \provisional{8.00} & \provisional{26.1} & \provisional{39.5} & \provisional{16.0} & \provisional{28.20} \\
\iconseed Seed 2.1 Pro & \provisional{8} & \provisional{229} & \provisional{2.50} & \provisional{24.7} & \provisional{36.7} & \provisional{16.3} & \provisional{21.01} \\
\iconglm GLM 5.2 & \provisional{5} & \provisional{361} & \provisional{2.06} & \provisional{20.4} & \provisional{35.0} & \provisional{12.5} & \provisional{24.40} \\
\icondeepseek DeepSeek V4 Pro & \provisional{0} & \provisional{150} & \provisional{0.90} & \provisional{1.8} & \provisional{23.9} & \provisional{1.5} & \provisional{13.04} \\
\bottomrule
\end{tabular}

%% file: content/C_ObservedTrajectories.tex
\FloatBarrier
\section{Claude-Opus-5.5 Trajectories: Measurement Details}
\label{sec:appendix-observed}

\paragraph{Trajectory archive.}
The analysis covers the \provisional{101}-task release cohort.
Behavioral measurements come from the available Claude-Opus-5.5
\textsc{CoWork} trajectories. Table~\ref{tab:partial-models}
reports benchmark outcomes evaluated against the complete test
contract. Each trajectory record includes its task version,
package hash, run ID, requested models, scaffold version,
runtime revision and budget. Runs resumed after a provider
interruption are treated as single observations.


\paragraph{Linking tests to requirements.}
For each F2P test node represented in the trajectory archive,
we map the tested requirements by comparing the per-node
requirement files, test source code and upstream change
titles with the \textsc{Oracle-Spec} for the same package.

\paragraph{Operational loss stages.}
We use card identifiers, document retrieval records and
disclosure events to identify where requirements were
missed or lost during information acquisition and use.

\paragraph{Disclosure and visibility.}
The workplace service logs each requirement disclosed
to the agent and links the disclosure to the question
that triggered it. A disclosure is considered visible
when its reply body appears in a tool result saved in
the trajectory. We associate questions with cards using
the card identifiers mentioned in the questions.
If the agent contacted a requirement's owner but never
mentioned its card, we classify the requirement as
\emph{owner contacted, card not raised}.

\paragraph{Behavior annotation.}
The trajectory reviewer, GPT-6~Astra at xhigh reasoning effort,
assesses observable actions and outcomes along ten strategy
dimensions and six collaboration dimensions. It evaluates ordered
trajectories before private grading outcomes are attached.
Each dimension uses four ordered categories:
\emph{ineffective} (an identifiable need remains
unaddressed), \emph{partial} (useful actions leave
identifiable gaps), \emph{effective} (appropriate actions
resolve the local need) and \emph{adaptive} (the agent
connects evidence across steps to resolve uncertainty
and adjust or verify its actions).

The categories are encoded as 0--3 for rank
analysis. Each dimension is assessed using its own
behavioral criteria. When the trajectory provides
insufficient evidence or no opportunity to exhibit
a behavior, the dimension is marked
\emph{not assessable}.

Category percentages are calculated separately for
each dimension using assessable runs. Spearman
correlations use runs with both an assessable category
and an observed per-task F2P score. A trajectory may
contribute to one dimension or diagnostic even when
another dimension lacks a completed assessment.

For owner-question follow-through, a relevant question
that lacks necessary follow-up may be classified as
partial. Following the question through to an actionable
answer is effective. Resolving consequential ambiguity
through linked questions and corroborating evidence
is adaptive.

\begin{table}[!htbp]
\centering
\small
\setlength{\tabcolsep}{5pt}
\input{tables/table_observed_dimensions}
\caption{Behavior category counts for Claude-Opus-5.5
trajectories in the \provisional{101}-task release
cohort. I: ineffective; P: partial; E: effective;
A: adaptive. Each dimension uses its own behavioral
criteria. Counts are reported separately by dimension.}
\label{tab:observed-dimensions}
\end{table}

\paragraph{Deferrals and rules.}
A deferred thread is an owner--card pair for which
the owner's first reply postpones the answer.
A follow-up is recorded when the agent later messages
the same owner about the same card, or when the
environment records a qualifying follow-up for that
thread.

Remaining budget is calculated from the eight-hour
limit and elapsed time since the first saved action.
Subsequent working time is measured through the
final saved action.

For runs whose brief requires test preservation
without a migration exception, we assess compliance
by counting lines changed or removed from existing
test source files in the submitted patch.

%% file: tables/table_observed_dimensions.tex
\begin{tabular}{lrrrrr}
\toprule
Dimension & Assessed $n$ & I & P & E & A \\
\midrule
Search strategy & 101 & 0 & 6 & 90 & 5 \\
Decomposition/prioritization & 101 & 0 & 17 & 75 & 9 \\
Hypothesis testing & 101 & 0 & 2 & 33 & 66 \\
Information value & 101 & 0 & 3 & 85 & 13 \\
Abstraction/transfer & 101 & 0 & 2 & 57 & 42 \\
Causal debugging & 101 & 0 & 2 & 28 & 71 \\
Feedback adaptation & 101 & 0 & 4 & 28 & 69 \\
Information integration & 101 & 0 & 18 & 77 & 6 \\
Verification design & 101 & 0 & 11 & 60 & 30 \\
Calibration/stopping & 101 & 0 & 62 & 33 & 6 \\
Scope reconstruction & 101 & 0 & 5 & 86 & 10 \\
Provenance/version reconciliation & 101 & 0 & 13 & 66 & 22 \\
Owner questions/follow-through & 101 & 0 & 54 & 33 & 14 \\
Notification handling & 101 & 0 & 31 & 60 & 10 \\
Evidence to implementation & 101 & 0 & 3 & 49 & 49 \\
Delivery accountability & 101 & 1 & 44 & 52 & 4 \\
\bottomrule
\end{tabular}

%% file: main.bbl
\begin{thebibliography}{44}
\providecommand{\natexlab}[1]{#1}
\providecommand{\url}[1]{\texttt{#1}}
\expandafter\ifx\csname urlstyle\endcsname\relax
  \providecommand{\doi}[1]{doi: #1}\else
  \providecommand{\doi}{doi: \begingroup \urlstyle{rm}\Url}\fi

\bibitem[{Alibaba Cloud Model Studio}(2026)]{qwen38}
{Alibaba Cloud Model Studio}.
\newblock {Qwen3.8-Max} and {Qwen3.8-Flash} model documentation, 2026.
\newblock \url{https://www.alibabacloud.com/help/en/model-studio/qwen3-8-max}.
\newblock For {Qwen3.8-Flash}, see \url{https://www.alibabacloud.com/help/en/model-studio/qwen3-8-flash}. Accessed November 13, 2026.

\bibitem[{Anthropic}(2026)]{claudeopus55}
{Anthropic}.
\newblock Introducing {Claude Opus 5.5}, 2026.
\newblock \url{https://www.anthropic.com/news/claude-opus-5-5}.
\newblock Official model announcement. Accessed October 19, 2026.

\bibitem[{ByteDance Seed}(2026)]{seed21}
{ByteDance Seed}.
\newblock {Seed2.1} officially released: Advancing {AI} productivity, 2026.
\newblock \url{https://seed.bytedance.com/en/seed2_1}.
\newblock Official model page. Accessed October 19, 2026.

\bibitem[{Cursor}(2026)]{cursorbench2026}
{Cursor}.
\newblock {CursorBench 4.0}, September 2026.
\newblock \url{https://cursor.com/evals}.
\newblock Version 4.0, released September 10, 2026; accessed September 15, 2026.

\bibitem[Das et~al.(2026)Das, Paliwal, and Sunil]{specificrealswe2026}
Snagnik Das, Siddhant Paliwal, and Janak Sunil.
\newblock Introducing {Real-SWE}.
\newblock Specific Labs, September 2026.
\newblock \url{https://withspecific.com/blog/real-swe}.
\newblock Original release, September 2026; current page presents version 1.1. Accessed November 13, 2026.

\bibitem[{Datacurve}(2026)]{deepswev11}
{Datacurve}.
\newblock {DeepSWE} v1.1, June 2026.
\newblock \url{https://deepswe.datacurve.ai/blog/deepswe-v1-1}.
\newblock Updated execution and node-id scoring release; accessed September 15, 2026.

\bibitem[{DeepSeek AI}(2026{\natexlab{a}})]{deepseeksuperagent2026}
{DeepSeek AI}.
\newblock {dsh-superagent}: Towards a more autonomous agent system.
\newblock \url{https://github.com/deepseek-ai/dsh-superagent-prototype}, 2026{\natexlab{a}}.
\newblock Research prototype.

\bibitem[{DeepSeek AI}(2026{\natexlab{b}})]{deepseekv41flash}
{DeepSeek AI}.
\newblock {DeepSeek-V4.1-Flash} release, September 2026{\natexlab{b}}.
\newblock \url{https://api-docs.deepseek.com/news/news260910}.
\newblock Official release note, September 10, 2026.

\bibitem[{DeepSeek AI}(2026{\natexlab{c}})]{deepseekv4pro}
{DeepSeek AI}.
\newblock {DeepSeek-V4-Pro} {GA} release, August 2026{\natexlab{c}}.
\newblock \url{https://api-docs.deepseek.com/news/news260813}.
\newblock Official release note, August 13, 2026.

\bibitem[Deng et~al.(2025)Deng, Da, Pan, He, Ide, Garg, Lauffer, Park, Pasari, Rane, Sampath, Krishnan, Kundurthy, Hendryx, Wang, Bharadwaj, Holm, Aluri, Zhang, Jacobson, Liu, and Kenstler]{swebenchpro2025}
Xiang Deng, Jeff Da, Edwin Pan, Yannis~Yiming He, Charles Ide, Kanak Garg, Niklas Lauffer, Andrew Park, Nitin Pasari, Chetan Rane, Karmini Sampath, Maya Krishnan, Srivatsa Kundurthy, Sean Hendryx, Zifan Wang, Vijay Bharadwaj, Jeff Holm, Raja Aluri, Chen Bo~Calvin Zhang, Noah Jacobson, Bing Liu, and Brad Kenstler.
\newblock {SWE-Bench Pro}: Can ai agents solve long-horizon software engineering tasks?, 2025.
\newblock \url{https://arxiv.org/abs/2509.16941}.

\bibitem[Edwards and Schuster(2026)]{askorassume2026}
Nicholas Edwards and Sebastian Schuster.
\newblock Ask or assume? uncertainty-aware clarification-seeking in coding agents, 2026.
\newblock \url{https://arxiv.org/abs/2603.26233}.

\bibitem[Huang et~al.(2026)Huang, Lee, Tng, and Ge]{deepswe2026}
Wenqi Huang, Charley Lee, Leonard Tng, and Serena Ge.
\newblock {DeepSWE}: Measuring frontier coding agents on original, long-horizon engineering tasks, 2026.
\newblock \url{https://arxiv.org/abs/2607.07946}.

\bibitem[Jain(2026)]{cursorbenchmethod2026}
Naman Jain.
\newblock How we compare model quality in {Cursor}, March 2026.
\newblock \url{https://cursor.com/blog/cursorbench}.
\newblock March 11, 2026; accessed September 15, 2026.

\bibitem[Jimenez et~al.(2024)Jimenez, Yang, Wettig, Yao, Pei, Press, and Narasimhan]{jimenez2024swebench}
Carlos~E. Jimenez, John Yang, Alexander Wettig, Shunyu Yao, Kexin Pei, Ofir Press, and Karthik Narasimhan.
\newblock {SWE-bench}: Can language models resolve real-world github issues?
\newblock In \emph{International Conference on Learning Representations (ICLR)}, 2024.
\newblock \url{https://arxiv.org/abs/2310.06770}.

\bibitem[Kim et~al.(2026)Kim, Gwon, Kim, Shim, and Lee]{realswe2026}
Gyuhyeong Kim, Hyojung Gwon, Jeonghyeon Kim, Kyuhong Shim, and Sunjae Lee.
\newblock {RealSWE}: A compositional evaluation of coding agents under realistic user requests, August 2026.
\newblock \url{https://arxiv.org/abs/2608.27831v2}.
\newblock Version 2, August 31, 2026.

\bibitem[King and Flanigan(2026)]{dialogueswe2026}
Brendan King and Jeffrey Flanigan.
\newblock {Dialogue SWE-Bench}: A benchmark for dialogue-driven coding agents, 2026.
\newblock \url{https://arxiv.org/abs/2606.13995}.

\bibitem[Ko et~al.(2007)Ko, DeLine, and Venolia]{ko2007information}
Amy~J. Ko, Robert DeLine, and Gina Venolia.
\newblock Information needs in collocated software development teams.
\newblock In \emph{29th International Conference on Software Engineering (ICSE)}, pages 344--353, 2007.
\newblock \doi{10.1109/ICSE.2007.45}.

\bibitem[Kondra et~al.(2026)Kondra, Mhatre, Kumar, Chu, Ahmad, Nangia, Agarwal, Dasgupta, Sinha, Sridharan, Dave, Graham, Song, Rahul, Lim, Thangamuthu, Singh, Liu, Pour, Chen, and Mattern]{frontierswe2026}
Rishyanth Kondra, Sanket Mhatre, Akshit Kumar, Evan Chu, Bilal~Bakht Ahmad, Ayush Nangia, Rajan Agarwal, Arpan Dasgupta, Animesh Sinha, Bhuvanesh Sridharan, Krupa Dave, Brendan Graham, Guanyu Song, Anirudh Rahul, Wei~Hern Lim, Abishek Thangamuthu, Ramneet Singh, Danna Liu, Navid Pour, Calvin Chen, and Justus Mattern.
\newblock {FrontierSWE v2}.
\newblock Proximal Blog, 2026.
\newblock \url{https://www.frontierswe.com/blog/v2}.
\newblock Version 2, 34 tasks, accessed September 15, 2026.

\bibitem[LaToza et~al.(2006)LaToza, Venolia, and DeLine]{latoza2006mental}
Thomas~D. LaToza, Gina Venolia, and Robert DeLine.
\newblock Maintaining mental models: A study of developer work habits.
\newblock In \emph{Proceedings of the 28th International Conference on Software Engineering (ICSE)}, pages 492--501, 2006.
\newblock \doi{10.1145/1134285.1134355}.

\bibitem[Li et~al.(2026)Li, Deng, Qian, Jiang, Huang, Yang, and Peng]{activeswe2026}
Haobin Li, Ping Deng, Weizhong Qian, Liang Jiang, Zhenyu Huang, Mouxing Yang, and Xi~Peng.
\newblock {Active-SWE}: Benchmarking coding agents for proactive bug fixing without issue reports, 2026.
\newblock \url{https://arxiv.org/abs/2608.04682}.

\bibitem[Miserendino et~al.(2025)Miserendino, Wang, Patwardhan, and Heidecke]{miserendino2025swelancer}
Samuel Miserendino, Michele Wang, Tejal Patwardhan, and Johannes Heidecke.
\newblock {SWE-Lancer}: Can frontier {LLMs} earn \$1 million from real-world freelance software engineering?, 2025.
\newblock \url{https://arxiv.org/abs/2502.12115}.

\bibitem[{Moonshot AI}(2026)]{kimik3}
{Moonshot AI}.
\newblock {Kimi K3} is here: Our most capable model, July 2026.
\newblock \url{https://forum.moonshot.ai/t/kimi-k3-is-here-our-most-capable-model/480}.
\newblock Official model announcement, July 22, 2026.

\bibitem[{OpenAI}(2024)]{swebenchverified}
{OpenAI}.
\newblock Introducing {SWE-bench} verified, 2024.
\newblock \url{https://openai.com/index/introducing-swe-bench-verified/}.

\bibitem[{OpenAI}(2026)]{gpt6astra}
{OpenAI}.
\newblock {GPT-6 Astra} model, 2026.
\newblock \url{https://developers.openai.com/api/docs/models/gpt-6-astra}.
\newblock Official model documentation. Accessed October 8, 2026.

\bibitem[Peng et~al.(2026)Peng, Huang, Zhang, Xu, Xiao, Hong, Lo, Qiu, Cao, He, and Cao]{icaebench2026}
Zhongyuan Peng, Dan Huang, Chuyu Zhang, Caijun Xu, Changyi Xiao, Shibo Hong, David Lo, Lin Qiu, Xuezhi Cao, Jiyuan He, and Yixin Cao.
\newblock {ICAE-Bench}: Evaluating coding agents as interactive project builders, 2026.
\newblock \url{https://arxiv.org/abs/2607.21217}.

\bibitem[Pu et~al.(2026)Pu, Lee, Sehwag, Lee, Zhu, Maurya, Raghavendra, Xue, and Denton]{pu2026lhaw}
George Pu, Michael~S. Lee, Udari~Madhushani Sehwag, David~J. Lee, Bryan Zhu, Yash Maurya, Mohit Raghavendra, Yuan Xue, and Samuel~Marc Denton.
\newblock {LHAW}: Controllable underspecification for long-horizon tasks, 2026.
\newblock \url{https://arxiv.org/abs/2602.10525}.

\bibitem[Raghavendra et~al.(2026)Raghavendra, Dan, Calvo, He, Mols, Anand, McCollum, Arakelyan, Bharadwaj, Park, Da, Rezaei, Liu, Kenstler, and He]{sweatlas2026}
Mohit Raghavendra, Soham Dan, Miguel~Romero Calvo, Yannis~Yiming He, Johannes~Baptist Mols, Gautam Anand, Cole McCollum, Edgar Arakelyan, Vijay Bharadwaj, Andrew Park, Jeff Da, MohammadHossein Rezaei, Bing Liu, Brad Kenstler, and Yunzhong He.
\newblock {SWE Atlas}: Benchmarking coding agents beyond issue resolution, 2026.
\newblock \url{https://arxiv.org/abs/2605.08366}.

\bibitem[Seshadri et~al.(2026)Seshadri, Cahyawijaya, Odumakinde, Singh, and Goldfarb-Tarrant]{lostinsimulation2026}
Preethi Seshadri, Samuel Cahyawijaya, Ayomide Odumakinde, Sameer Singh, and Seraphina Goldfarb-Tarrant.
\newblock Lost in simulation: {LLM}-simulated users are unreliable proxies for human users in agentic evaluations, 2026.
\newblock \url{https://arxiv.org/abs/2601.17087}.

\bibitem[Shi et~al.(2026)Shi, Dhandhania, Narasimhan, and Barres]{shi2026hypertau}
Quan Shi, Keshav Dhandhania, Karthik Narasimhan, and Victor Barres.
\newblock {$\tau^{\tau}$}-bench: An environment for end-to-end, realistic agent construction, 2026.
\newblock \url{https://arxiv.org/abs/2609.04611}.

\bibitem[Sillito et~al.(2008)Sillito, Murphy, and Volder]{sillito2006questions}
Jonathan Sillito, Gail~C. Murphy, and Kris~De Volder.
\newblock Asking and answering questions during a programming change task.
\newblock \emph{IEEE Transactions on Software Engineering}, 34\penalty0 (4):\penalty0 434--451, 2008.
\newblock \doi{10.1109/TSE.2008.26}.

\bibitem[{The Anthropic Institute}(2026)]{anthropic2026recursive}
{The Anthropic Institute}.
\newblock When {AI} builds itself: Our progress toward recursive self-improvement, and its implications.
\newblock \url{https://www.anthropic.com/institute/recursive-self-improvement}, 2026.
\newblock Official report. Accessed October 2026.

\bibitem[{Volcano Engine}(2026)]{seedevolving}
{Volcano Engine}.
\newblock Model release announcements: {Doubao-Seed-Evolving}, 2026.
\newblock \url{https://docs.volcengine.com/docs/ark/model-release-announcement?lang=zh}.
\newblock Official release log with dated model updates. Accessed November 13, 2026.

\bibitem[Wang et~al.(2025)Wang, Li, Song, Xu, Tang, Zhuge, Pan, Song, Li, Singh, Tran, Li, Ma, Zheng, Qian, Shao, Muennighoff, Zhang, Hui, Lin, Brennan, Peng, Ji, and Neubig]{wang2025openhands}
Xingyao Wang, Boxuan Li, Yufan Song, Frank~F. Xu, Xiangru Tang, Mingchen Zhuge, Jiayi Pan, Yueqi Song, Bowen Li, Jaskirat Singh, Hoang~H. Tran, Fuqiang Li, Ren Ma, Mingzhang Zheng, Bill Qian, Yanjun Shao, Niklas Muennighoff, Yizhe Zhang, Binyuan Hui, Junyang Lin, Robert Brennan, Hao Peng, Heng Ji, and Graham Neubig.
\newblock {OpenHands}: An open platform for {AI} software developers as generalist agents.
\newblock In \emph{International Conference on Learning Representations (ICLR)}, 2025.
\newblock \url{https://arxiv.org/abs/2407.16741}.

\bibitem[Xia et~al.(2024)Xia, Deng, Dunn, and Zhang]{xia2024agentless}
Chunqiu~Steven Xia, Yinlin Deng, Soren Dunn, and Lingming Zhang.
\newblock Agentless: Demystifying {LLM}-based software engineering agents, 2024.
\newblock \url{https://arxiv.org/abs/2407.01489}.

\bibitem[Xia et~al.(2018)Xia, Bao, Lo, Xing, Hassan, and Li]{xia2018comprehension}
Xin Xia, Lingfeng Bao, David Lo, Zhenchang Xing, Ahmed~E. Hassan, and Shanping Li.
\newblock Measuring program comprehension: A large-scale field study with professionals.
\newblock \emph{IEEE Transactions on Software Engineering}, 44\penalty0 (10):\penalty0 951--976, 2018.
\newblock \doi{10.1109/TSE.2017.2734091}.

\bibitem[Xu et~al.(2025)Xu, Song, Li, Tang, Jain, Bao, Wang, Zhou, Guo, Cao, Yang, Lu, Martin, Su, Maben, Mehta, Chi, Jang, Xie, Zhou, and Neubig]{xu2024agentcompany}
Frank~F. Xu, Yufan Song, Boxuan Li, Yuxuan Tang, Kritanjali Jain, Mengxue Bao, Zora~Z. Wang, Xuhui Zhou, Zhitong Guo, Murong Cao, Mingyang Yang, Hao~Yang Lu, Amaad Martin, Zhe Su, Leander Maben, Raj Mehta, Wayne Chi, Lawrence Jang, Yiqing Xie, Shuyan Zhou, and Graham Neubig.
\newblock {TheAgentCompany}: Benchmarking {LLM} agents on consequential real world tasks.
\newblock In \emph{Advances in Neural Information Processing Systems (NeurIPS), Datasets and Benchmarks Track}, 2025.
\newblock \url{https://arxiv.org/abs/2412.14161}.

\bibitem[Yang et~al.(2024)Yang, Jimenez, Wettig, Lieret, Yao, Narasimhan, and Press]{yang2024sweagent}
John Yang, Carlos~E. Jimenez, Alexander Wettig, Kilian Lieret, Shunyu Yao, Karthik Narasimhan, and Ofir Press.
\newblock {SWE-agent}: Agent-computer interfaces enable automated software engineering.
\newblock In \emph{Advances in Neural Information Processing Systems (NeurIPS)}, volume~37, pages 50528--50652, 2024.
\newblock \url{https://arxiv.org/abs/2405.15793}.

\bibitem[Yang et~al.(2025{\natexlab{a}})Yang, Jimenez, Zhang, Lieret, Yang, Wu, Press, Muennighoff, Synnaeve, Narasimhan, Yang, Wang, and Press]{yang2024swebenchmm}
John Yang, Carlos~E. Jimenez, Alex~L. Zhang, Kilian Lieret, Joyce Yang, Xindi Wu, Ori Press, Niklas Muennighoff, Gabriel Synnaeve, Karthik~R. Narasimhan, Diyi Yang, Sida~I. Wang, and Ofir Press.
\newblock {SWE-bench} multimodal: Do ai systems generalize to visual software domains?
\newblock In \emph{International Conference on Learning Representations (ICLR)}, 2025{\natexlab{a}}.
\newblock \url{https://arxiv.org/abs/2410.03859}.

\bibitem[Yang et~al.(2025{\natexlab{b}})Yang, Lieret, Jimenez, Wettig, Khandpur, Zhang, Hui, Press, Schmidt, and Yang]{yang2025swesmith}
John Yang, Kilian Lieret, Carlos~E. Jimenez, Alexander Wettig, Kabir Khandpur, Yanzhe Zhang, Binyuan Hui, Ofir Press, Ludwig Schmidt, and Diyi Yang.
\newblock {SWE-smith}: Scaling data for software engineering agents, 2025{\natexlab{b}}.
\newblock \url{https://arxiv.org/abs/2504.21798}.
\newblock Introduces SWE-bench Multilingual (300 instances, 42 repositories, 9 languages).

\bibitem[Yao et~al.(2024)Yao, Shinn, Razavi, and Narasimhan]{yao2024taubench}
Shunyu Yao, Noah Shinn, Pedram Razavi, and Karthik Narasimhan.
\newblock $\tau$-bench: A benchmark for tool-agent-user interaction in real-world domains, 2024.
\newblock \url{https://arxiv.org/abs/2406.12045}.

\bibitem[{Z.ai}(2026{\natexlab{a}})]{glm52}
{Z.ai}.
\newblock {GLM-5.2} overview, 2026{\natexlab{a}}.
\newblock \url{https://docs.z.ai/guides/llm/glm-5.2}.
\newblock Official developer documentation. Accessed October 19, 2026.

\bibitem[{Z.ai}(2026{\natexlab{b}})]{glm53}
{Z.ai}.
\newblock {GLM-5.3} overview, 2026{\natexlab{b}}.
\newblock \url{https://docs.z.ai/guides/llm/glm-5.3}.
\newblock Official developer documentation. Accessed October 19, 2026.

\bibitem[Zan et~al.(2025)Zan, Huang, Liu, Chen, Zhang, Xin, Chen, Liu, Zhong, Li, Liu, Xiao, Chen, Zhang, Su, Liu, Long, Shen, and Xiang]{zan2025multiswebench}
Daoguang Zan, Zhirong Huang, Wei Liu, Hanwu Chen, Linhao Zhang, Shulin Xin, Lu~Chen, Qi~Liu, Xiaojian Zhong, Aoyan Li, Siyao Liu, Yongsheng Xiao, Liangqiang Chen, Yuyu Zhang, Jing Su, Tianyu Liu, Rui Long, Kai Shen, and Liang Xiang.
\newblock {Multi-SWE-bench}: A multilingual benchmark for issue resolving, 2025.
\newblock \url{https://arxiv.org/abs/2504.02605}.

\bibitem[Zhou et~al.(2026)Zhou, Chong, Kim, Peng, Shu, Wu, Han, Yuan, Zhuang, Kim, Ju, Ju, Yoon, and Lo]{swerpg2026}
Xin Zhou, Chun~Yong Chong, Kisub Kim, Yun Peng, Rui Shu, Zihan Wu, Xu~Han, Guowen Yuan, Zeyang Zhuang, Jounghoon Kim, Jeongjin Ju, Seongmin Ju, Taein Yoon, and David Lo.
\newblock A unified issue resolution benchmark for requirement clarification, planning, and code generation for coding agents, 2026.
\newblock \url{https://arxiv.org/abs/2608.09072}.

\end{thebibliography}
